\documentclass[preprint,nofootinbib,aps,superscriptaddress,eqsecnum]{revtex4-1}
\pdfoutput=1
\usepackage{graphicx}
\usepackage[dvipsnames]{xcolor}
\usepackage{mathrsfs}
\usepackage{array}
\usepackage{epstopdf}
\usepackage{tikz}

\DeclareGraphicsExtensions{.eps}
\usepackage{soul}
\usepackage{amssymb,amsmath,subfigure,slashed}
\usepackage[utf8]{inputenc}

\newcommand{\lapprox}{%
\mathrel{%
\setbox0=\hbox{$<$}
\raise0.6ex\copy0\kern-\wd0
\lower0.65ex\hbox{$\sim$}
}}
\newcommand{\gapprox}{%
\mathrel{%
\setbox0=\hbox{$>$}
\raise0.6ex\copy0\kern-\wd0
\lower0.65ex\hbox{$\sim$}
}}
\newcolumntype{P}[1]{>{\centering\arraybackslash}p{#1}}
\newcommand{\ba}{\begin{array}}
\newcommand{\ea}{\end{array}}
\newcommand{\bd}{\begin{displaymath}}
\newcommand{\ed}{\end{displaymath}}
\newcommand{\beq}{\begin{equation}}
\newcommand{\eeq}{\end{equation}}
\newcommand{\bea}{\begin{eqnarray}}
\newcommand{\eea}{\end{eqnarray}}

\newcommand{\ra}{\rightarrow}
\newcommand{\nn}{\nonumber}
\def\ie{ {\em i.e.,\ }}

\def\bra{\langle}
\def\ket{\rangle}

\def\a{\alpha}

\def\b{\beta}
\def\g{\gamma}

\def\m{\mu}
\def\n{\nu}

\def\q2 {q^2}

\def\bt{\begin{table}}
\def\et{\end{table}}

\catcode`@=11 
\def \gsim{\mathrel{\mathpalette\@versim>}}
\def \lsim{\mathrel{\mathpalette\@versim<}}
\def \@versim#1#2{\lower0.4ex\vbox{\baselineskip\z@skip\lineskip\z@skip
     \lineskiplimit\z@\ialign{$\m@th#1\hfil##\hfil$%
     \crcr#2\crcr\sim\crcr}}}
\catcode`@=12 

\begin{document}

\renewcommand*{\thefootnote}{\fnsymbol{footnote}}

\begin{center}


{\large\bf Interactions of Astrophysical Neutrinos with Dark Matter: A model building perspective }
\vskip 15pt
Sujata Pandey\footnote{E-mail: phd1501151007@iiti.ac.in},
Siddhartha Karmakar\footnote{E-mail: phd1401251010@iiti.ac.in} and Subhendu
Rakshit\footnote{E-mail: rakshit@iiti.ac.in} 
\\[2mm]

{\em Discipline of Physics, Indian Institute of Technology Indore,\\
 Khandwa Road, Simrol, Indore - 453\,552, India}
\\[20mm]
\end{center}

\begin{abstract} 
\vskip 20pt

We explore the possibility that high energy astrophysical neutrinos can interact with the dark matter on their way to Earth. Keeping in mind that new physics might leave its signature at such energies, we have considered all possible topologies for effective interactions between neutrino and dark matter. Building models, that give rise to a significant flux suppression of astrophysical neutrinos at Earth, is rather difficult. We present a $Z^{\prime}$-mediated model in this context. Encompassing a large variety of models, a wide range of dark matter masses from $10^{-21}$~eV up to a TeV, this study aims at highlighting the challenges one encounters in such a model building endeavour after satisfying various cosmological constraints, collider search limits and electroweak precision measurements. 
 
 \end{abstract}

\vskip 1 true cm
\maketitle

\setcounter{footnote}{0}
\renewcommand*{\thefootnote}{\arabic{footnote}}

\section{Introduction}
IceCube has been to designed to detect high energy astrophysical neutrinos of extragalactic origin. Beyond neutrino energies of $\sim 20$~TeV the background of atmospheric neutrinos get diminished and the neutrinos of higher energies are attributed to extragalactic sources~\cite{Denton:2017csz}. However, there is a paucity of high energy neutrino events observed at IceCube for neutrino energies greater than $\sim 400$~TeV~\cite{Aartsen:2017mau}. There are a few events around $\sim 1$~PeV or higher, whose origin perhaps can be described by the decay or annihilation of very heavy new particles~\cite{Boucenna:2015tra,Borah:2017xgm,Zavala:2014dla,Dev:2016qeb,Hiroshima:2017hmy,Lambiase:2018yql,Murase:2015gea,Dhuria:2017ihq} or even without the help of any new physics~\cite{Murase:2016gly,Chen:2013dza,Murase:2015ndr}. In the framework of standard astrophysics, high energy cosmic rays of energies up to $10^{20}$~eV have been observed, which leads to the prediction of the existence of neutrinos of such high energies as well~\cite{Aab:2017tyv,Bahcall:1999yr,Waxman:1998yy}. 
In this context, it is worth exploring whether the flux of such neutrinos can get altered due to their interactions with DM particles. However, it is challenging to build such models given the relic abundance of dark matter. Few such attempts have been made in literature but these models also suffer from cosmological and collider constraints. Hence, in this paper, we take a model building perspective to encompass a large canvas of such interactions that can lead to appreciable flux suppression at IceCube.

In presence of neutrino-DM interaction, the flux of astrophysical neutrinos passing through isotropic DM background is attenuated by a factor $\sim \exp(-n \sigma L)$. Here $n$ denotes number density of DM particles, $L$ is the distance traversed by the neutrinos in the DM  background and $\sigma$ represents the cross-section of neutrino-DM interaction. 
The neutrino-DM interaction can produce appreciable flux suppression only when the number of interactions given by 
$n \sigma L$ is $\gtrsim \mathcal{O}(1)$.
For lower masses of DM, the number density is significant. 
But the cross-section depends on both the structure of the neutrino-DM interaction vertex and the DM mass.  
The neutrino-DM cross-section might increase with DM mass for some particular interactions.
Hence, it is essentially the interplay between DM number density and the nature of the neutrino-DM interaction, which determines whether a model leads to a significant flux suppression. 
As a pre-filter to identify such cases we impose the criteria that the interactions must lead to at least $1\%$ suppression of the incoming neutrino flux.
For the rest of the paper, a flux suppression of less than $1\%$ has been addressed as `not significant'.
 While checking an interaction against this criteria, we consider the entire energy range of the astrophysical neutrinos. 
If an interaction leads to $1\%$ change in neutrino flux after considering the relevant collider and cosmological constraints in any part of this entire energy range, it passes this empirical criteria.
We explore a large range of DM mass ranging from sub-eV regimes to WIMP scenarios. 
In the case of sub-eV DM, we investigate the ultralight scalar DM which can exist as a Bose-Einstein condensate in the present Universe. 

In general, various aspects of the neutrino-DM interactions have been addressed in the literature~\cite{Boehm:2013jpa,Campo:2017nwh,Wilkinson:2014ksa,Escudero:2015yka,Barranco:2010xt,Reynoso:2016hjr,Arguelles:2017atb,deSalas:2016svi,Huang:2018cwo}. 
The interaction of astrophysical neutrinos with cosmic neutrino background can lead to a change in the flux of such neutrinos as well~\cite{Ng:2014pca,DiFranzo:2015qea,Araki:2015mya,Mohanty:2018cmq,Chauhan:2018dkd,Kelly:2018tyg,Cherry:2016jol,Shoemaker:2015qul,Ibe:2014pja}. 
But it is possible that the dark matter number density is quite large compared to the number density of the relic neutrinos, leading to more suppression of the astrophysical neutrino flux. 

To explore large categories of models with neutrino-DM interactions, we take into account the renormalisable as well as the non-renormalisable models.
In case of non-renormalisable models, we consider neutrino-DM effective interactions up to dimension-eight.
However, it is noteworthy that for a wide range of DM mass the centre-of-mass energy of the neutrino-DM scattering can be such that the effective interaction scale can be considered to be as low as $\sim 10$~MeV. We discuss relevant collider constraints on both the effective interactions and renormalisable models. We consider thermal DM candidates with masses ranging in MeV$-$TeV range as well as non-thermal ultralight DM with sub-eV masses. For the thermal DM candidates, we demonstrate the interplay between constraints from relic density, collisional damping and  the effective number of light neutrinos on the respective parameter space. 
Only for a few types of interactions, one can obtain  significant flux suppressions.  
For the renormalisable interaction leading to flux suppression, we present a UV-complete model taking into account anomaly cancellation, collider constraints and precision bounds.

In Sec.~\ref{DMC} we discuss the nature of the DM candidates that might lead to flux suppression of neutrinos. 
In Sec.~\ref{EFT} we present the non-renormalisable models,\ie the effective neutrino-DM  interactions categorised into four topologies. 
In Sec.~\ref{renmodels} we present three renormalisable neutrino-DM interactions and the corresponding cross-sections in case of thermal as well as non-thermal ultralight scalar DM. In Sec.~\ref{zprime} we present a UV-complete model mediated by a light $Z'$ which leads to a significant flux suppression. Finally in Sec.~\ref{conclusion} we summarise our key findings and eventually conclude.

\section{Dark Matter Candidates}
\label{DMC}

In this section, we systematically narrow down the set of DM candidates we are interested in considering a few cosmological and phenomenological arguments.

 The Lambda cold dark matter~($\Lambda$CDM) model explains the anisotropies of cosmic microwave background~(CMB) quite well. 
The weakly interacting massive particles~(WIMP) are interesting candidates of CDM, mostly because they appear in well-motivated BSM theories of particle physics. 
Nevertheless, CDM with sub-GeV masses are also allowed. The most stringent lower bound on the mass of CDM comes from the effective number of neutrinos~($N_{\rm eff}$) implied by the CMB measurements from the Planck satellite.  For complex and real scalar DM as well as Dirac and Majorana fermion DM, this lower bound comes out to be $\sim 10$~MeV~\cite{Boehm:2013jpa,Campo:2017nwh}.
Thermal DM with masses lower than $\sim 10$~MeV are considered hot and warm DM candidates and are allowed to make up only a negligible fraction of the total dark matter abundance~\cite{Bode:2000gq}. 
 The ultralight non-thermal Bose-Einstein condensate~(BEC) dark matter with mass $\sim 10^{-21} - 1$~eV is also a viable cold dark matter candidate~\cite{Boehm:2003hm}. In the rest of this paper, unless mentioned otherwise, by ultralight DM we refer to the non-thermal ultralight BEC DM.

Numerical simulations with the $\Lambda$CDM model show a few tensions with cosmological observations at small,\ie galactic scales~\cite{Salucci:2002nc,Tasitsiomi:2002hi,Blok:2002tr}. It predicts too many sub-halos of DM in the vicinity of a galactic DM halo, thus predicting the existence of many satellite galaxies which have not been observed. This is known as the missing satellite problem~\cite{Klypin:1999uc}. It also predicts a `cusp' nature in the galactic rotational curves,\ie a density profile that is proportional to $r^{-1}$ near the centre, with $r$ being the radial distance from the centre of a galaxy. On the contrary, the observed rotational curves show a `core',\ie a constant nature. This is known as the cusp/core problem~\cite{Navarro:1995iw}. Ultralight scalar DM provides an explanation to such small-scale cosmological problems. In such models, at small scales, the quantum pressure of ultralight bosons prevent the overproduction of sub-halos and dwarf satellite galaxies~\cite{Alcubierre:2001ea,Hu:2000ke,Harko:2011jy}. Also, choosing suitable boundary condition while solving the Schr{\"o}dinger equation for the evolution of ultralight DM wavefunction can alleviate the cusp/core problem~\cite{Hu:2000ke,Peebles:2000yy,Matos:2007zza,Su:2010bj}, making ultralight scalar an interesting, even preferable alternative to WIMP.   
Ultralight DM form BEC at an early epoch and acts like a ``cold" species in spite of their tiny masses~\cite{Sikivie:2009qn}.
Numerous searches of these kinds for DM are underway, namely ADMX~\cite{Duffy:2006aa}, CARRACK~\cite{Tada:1999tu} etc.
It has been recently proposed that gravitational waves can serve as a probe of ultralight BEC DM as well~\cite{Dev:2016hxv}.
But the ultralight fermionic dark matter is not a viable candidate for CDM, because it can not form such a condensate and is, therefore ``hot". The case of ultralight vector dark matter also has been studied in the literature~\cite{Graham:2015rva}.

The scalar DM can transform under $SU(2)_L$ as a part of any multiplet. In the case of a doublet or higher representations, the DM candidate along with other degrees of freedom in the dark sector couple with $W^{\pm}, Z$ bosons at the tree level. This leads to stringent bounds on their masses as light DM candidates can heavily contribute to the decay width of SM gauge bosons, and hence, are ruled out from the precision experiments. 
Moreover, Higgs-portal WIMP DM candidates with $m_{\rm DM} \ll m_h/2$ are strongly constrained from the Higgs invisible decay width as well.
The failure of detecting DM particles in collider searches and the direct DM detection experiments rule out a vast range of parameter space for  WIMPs. 
In light of current LUX and XENON data, amongst low WIMP DM masses, only a narrow mass range near the Higgs funnel region,\ie $m_{\rm DM} \sim 62$~GeV, survives~\cite{Aprile:2015uzo,Akerib:2016vxi,Athron:2017kgt}. 
As alluded to earlier, the ultralight scalar DM can transform only as a singlet  under $SU(2)_L$ because of its tiny mass.

We investigate the scenarios of scalar dark matter, both thermal and ultralight, as possible candidates to cause flux suppression of the high energy astrophysical neutrinos.  
Such a suppression depends on the length of the path the neutrino travels in the isotropic DM background and the mean free path of neutrinos, which depends on the cross-section of neutrino-DM interaction and the number density of DM particles. 
We take the length traversed by neutrinos to be $\sim 200$~Mpc, the distance from the nearest group of quasars~\cite{Schawinski:2010up}, which yields a conservative estimate for the flux suppression. Moreover, we consider the density of the isotropic DM background to be $\sim 1.2 \times 10^{-6}~\text{GeV cm}^{-3}$~\cite{Agashe:2014kda}.  
Comparably, in the case of WIMP DM, the number density is much smaller, making it interesting to investigate whether the cross-section of neutrino-DM interaction in these cases can be large enough to compensate for the smallness of DM number density. This issue will be addressed in a greater detail in Sec.~IV.

\section{Effective Interactions}
\label{EFT}
\begin{figure}[h!]
 \begin{center}
\subfigure[]{
 \includegraphics[width=0.9in,height=1.2in, angle=0]{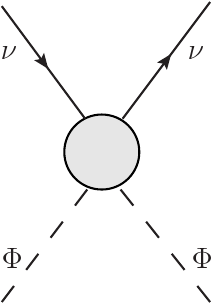}}
 \hskip 35pt
 \subfigure[]{
 \includegraphics[width=0.85in,height=1.2in, angle=0]{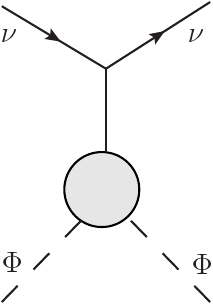}}\\
 \subfigure[]{
 \includegraphics[width=0.9in,height=1.2in, angle=0]{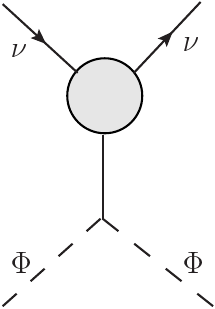}}
 \hskip 35pt
 \subfigure[]{
 \includegraphics[width=0.9in,height=1.2in, angle=0]{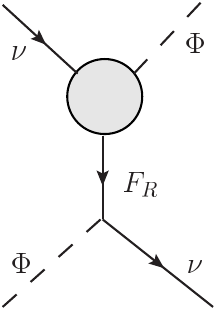}}
 \caption{Topologies of effective neutrino-DM interactions. Fig.~(a), (b), (c) and (d) represent topology I, II, III and IV respectively.}
 \label{topology}
\end{center} 
 \end{figure}

In order to exhaust the set of higher dimensional effective interactions contributing to the process of neutrino scattering off scalar DM particles, we consider four topologies of diagrams representing all the possibilities as depicted in fig.~\ref{topology}. Topology~I represents a contact type of interaction.
In case of topologies~II, III, and IV we consider higher dimensional interaction in one of the vertices while the neutrino-DM interaction is mediated by either a vector, a scalar or a fermion, whenever appropriate.

$\nu {\bar\nu}$\,DM\,DM effective interactions can arise from  higher dimensional gauge-invariant interactions as well. 
 In this case, the bounds on such interactions may be more restrictive than the case where the mediators are light and hence, are parts of the low energy spectrum. In general low energy neutrino-DM effective interactions need not reflect explicit gauge invariance.

We discuss the bounds on the effective interactions based on LEP monophoton searches and the measurement of the $Z$ decay width. The details of our implementation of these two bounds are as follows:  

\vskip 10pt
\noindent$\bullet$ {\bf Bounds from LEP monophoton searches} 

For explicitly gauge-invariant effective interactions, $\nu {\bar\nu}$\,DM\,DM interactions come along with $l^{+} l^{-}$\,DM\,DM interactions. $e^+ {e^-}$\,DM\,DM interactions can be constrained from the channel $e^{+}e^{-} \rightarrow \g + \slashed{E_{T}}$ using FEMC data at DELPHI detector in LEP for $190$ GeV $\leq \sqrt{s}\leq 209$ GeV. To extract a conservative estimate on the interaction, we assume that the new contribution saturates the error in the measurement of the cross-section $1.71 \pm 0.14 $ pb at 1$\sigma$~\cite{Abdallah:2003np}. By the same token, we consider only one effective interaction at a time. 
 The corresponding kinematic cuts on the photon at the final state were imposed in accordance with the FEMC detector: 
$20.4 \leq E_{\gamma} \,{\rm(in~GeV)} \leq 91.8$, $12^{\circ} \leq \theta_{\gamma} \leq 32^{\circ}$ and $148^{\circ} \leq \theta_{\gamma} \leq 168^{\circ}$. Here $E_{\gamma}$ stands for the energy of the outgoing photon and $\theta_{\gamma}$ is its angle with the beam axis. We use \texttt{FeynRules-2.3.32}~\cite{Christensen:2008py}, \texttt{CalcHEP-3.6.27}~\cite{Belyaev:2012qa} and \texttt{MadGraph-2.6.1}~\cite{Alwall:2014hca} for computations.

Although here we considered gauge-invariant interactions, $\nu {\bar\nu}$\,DM\,DM interactions can be directly constrained from the monophoton searches due to the existence of the channel $e^{+}e^{-} \rightarrow \gamma\nu\bar{\nu}$\,DM\,DM \textit{via} a $Z$ boson.
But such bounds are generally weaker than the bounds obtained from $Z$ decay which we are going to consider next.

$\mu^+ {\mu^-}$\,DM\,DM interactions can contribute to the muon decay width which is measured with an error of $10^{-4}$\%. However, the partial decay width of the muon \textit{via} $\mu\ra \nu_{\mu} e^- \bar{\nu}_e$\,DM\,DM channel is negligible compared to the error. Hence, these interactions are essentially unbounded from such considerations. The percentage error in the decay width for tauon is even larger and hence, the same is true for $\tau^+ {\tau^-}$\,DM\,DM interactions.

\vskip 10pt
\noindent$\bullet$ {\bf Bounds from the leptonic decay modes of the $Z$-boson}

The effective $\nu {\bar\nu}$\,DM\,DM interactions can be constrained from 
 the invisible decay width of the $Z$ boson which is measured to be $ \Gamma(Z \ra inv) =0.48\pm 0.0015$~GeV~\cite{Agashe:2014kda}. 
When the gauge-invariant forms of such effective interactions are taken into account, $l^{+} l^{-}$\,DM\,DM interactions may be constrained from the experimental error in the partial decay width of the channel $Z\ra l^{+} l^{-}$: $\Delta \Gamma (Z\ra l^{+} l^{-}) \sim 0.176, 0.256, 0.276$~MeV for $\ell=e, \mu, \tau$ at 1$\sigma$~\cite{Agashe:2014kda}. To extract conservative upper limits on the strength of such interactions, one can saturate this error with the partial decay width  $\Gamma(Z\ra l^{+} l^{-}$\,DM\,DM).

 If such interactions are mediated by some  particle, say a  light $Z^\prime$, then a stringent bound can be obtained by saturating $\Delta \Gamma (Z\ra l^{+} l^{-})$ with $\Gamma(Z\ra l^{+} l^{-} Z^\prime)$. Similar considerations hold true for $Z\ra \nu {\bar \nu}$\,DM\,DM mediated by a $Z^\prime$. We note in passing that such constraints from $Z$ decay measurements are particularly interesting for light DM candidates.

\subsection{Topology I}
\label{Topa}
In this subsection effective interactions up to dimension 8 have been considered which can give rise to neutrino-DM scattering. The phase space factor for the interaction of the high energy neutrinos with DM can be found in appendix~\ref{appendixA1}.

\begin{enumerate}

\item 
A six-dimensional interaction term leading to neutrino-DM scattering can be written as, 
\bea
\mathcal{L} \supset  \frac{c^{(1)}_{l}}{\Lambda^2}(\bar{\nu} i \slashed{\partial} \nu) (\Phi^* \Phi),
\label{op1}
\eea
where $\nu$ is SM neutrino, $\Phi$ is the scalar DM and $\Lambda$ is the effective interaction scale.

Now, for this interaction, the constraint from $Z$ invisible decay reads $c_{l}^{(1)}/\Lambda^2 \lsim 8.8 \times 10^{-3}$~GeV$^{-2}$. The bounds from the measurements of the channel $Z \rightarrow l^{+} l^{-}$ are dependent on the lepton flavours, and are found to be: $c^{(1)}_{e}/\Lambda^2 \lesssim 5.0\times10^{-3}$~GeV$^{-2}$, $c^{(1)}_{\mu}/\Lambda^2 \lesssim 6.0\times10^{-3}$~GeV$^{-2}$ and $c^{(1)}_{\tau}/\Lambda^2 \lesssim 6.2\times10^{-3}$~GeV$^{-2}$. 
The gauge-invariant form of this effective interaction leads to a five-point vertex of $\nu \bar{\n} \Phi \Phi Z$, which in turn leads to a new four-body decay channel of the $Z$ boson. Due to the existence of such a vertex, the bound on this interaction from the $Z$ decay width reads $c_{l}^{(1)}/\Lambda^2 \lesssim 9 \times 10^{-3}$~GeV$^{-2}$.
The electron-DM effective interactions can be further constrained from the measurements of $e^{+}e^{-} \rightarrow \g + \slashed{E_{T}}$, leading to $c^{(1)}_{e}/\Lambda^2 \lesssim 10^{-4}$~GeV$^{-2}$.
It can be seen that for the effective interaction with electrons, the bound from the measurement of the cross-section in the channel $e^{+}e^{-} \rightarrow \g + \slashed{E_{T}}$ can be quite stringent even compared to the bounds coming from the $Z$ decay width. 
Among all the constraints pertaining to such different considerations, if one assumes the least stringent bound, the interaction still leads to only $\lesssim 1\%$ flux suppression.
The renormalisable model discussed in Sec.~\ref{ren_ferm} is one of the  scenarios that leads to the effective interaction as in eq.~(\ref{op1}).

\item
Another six-dimensional interaction is given as:
 \bea
\mathcal{L} \supset  \frac{c^{(2)}_{l}}{\Lambda^{2}} (\bar{\nu} \gamma^\mu \nu)(\Phi^* \partial_\mu \Phi- \Phi \partial_\mu \Phi^*).
\label{op2}
\eea

The constraint from the measurement of the decay width in the $Z \rightarrow inv$ channel reads $c_{l}^{(2)}/\Lambda^2 \lsim 1.8 \times 10^{-2}$~GeV$^{-2}$ for light DM. 
The bounds on the gauge-invariant form of the interaction in eq.~(\ref{op2}) from the measurement of $Z \rightarrow l^{+} l^{-}$ reads $c_{e}^{(2)}/\Lambda^2 \lsim 1.7 \times 10^{-2}$~GeV$^{-2}$, $c_{\m}^{(2)}/\Lambda^2 \lsim 1.2 \times 10^{-2}$~GeV$^{-2}$ and $c_{\tau}^{(2)}/\Lambda^2 \lsim 1.3 \times 10^{-2}$~GeV$^{-2}$.
The bound from the channel $e^{+}e^{-} \rightarrow \g + \slashed{E_{T}}$ reads $c^{(2)}_{e}/\Lambda^{2} \lesssim 2.6 \times 10^{-5}$~GeV$^{-2}$.
Even with the value $c_{l}^{(2)}/\Lambda^2 \sim 10^{-2}$~GeV$^{-2}$, such an effective interaction does not give rise to an appreciable flux suppression due to the structure of the vertex.

\item
\label{sub} Another five dimensional effective Lagrangian for the neutrino-DM four-point interaction  is given by:
\bea
\mathcal{L} \supset \frac{c_{l}^{(3)}}{\Lambda} \bar{\nu^{c}} \nu \,\, \Phi^{\star} \Phi.
\label{op3} 
\eea

The above interaction gives rise to neutrino mass at the  loop-level which is proportional to $m_{\text{DM}}^2$. This, in turn, leads to a bound on the effective interaction due to the smallness of neutrino mass, 
\beq
\frac{c^{(3)}_{l}}{\Lambda} \lesssim 16 \pi^2 \frac{m_{\nu}}{m_{\text{DM}}^{2}} \sim 1.6 \pi^2 \Big(\frac{1\, \text{eV}}{m_{\text{DM}}}\Big)^2 \Big(\frac{m_{\nu}}{0.1 \, \text{eV}} \Big)\text{eV}^{-1},
\label{mnu}
\eeq
up to a factor of $\mathcal{O}(1)$. In the ultralight regime mass of DM $\lesssim 1$ eV. Hence eq.~(\ref{mnu}) does not lead to any useful constraint on $c^{(3)}_{l}/\Lambda$.
The constraint from invisible $Z$ decay on this interaction reads $c^{(3)}_{l}/{\Lambda} \leq 0.5$ GeV$^{-1}$, independent of neutrino flavour. 
The gauge-invariant form of this interaction does not contain additional vertices involving the charged leptons and hence leads to no further constraints. 
For such a value of coupling, there can be a significant flux suppression for the entire range of ultralight DM mass, independent of the energy of the  incoming neutrino as shown in fig.~\ref{fig:mesh}.

In passing, we note that the interaction can be written in a gauge-invariant manner at the tree-level only when $\Delta$, a $SU(2)_L$ triplet with hypercharge $Y=2$, is introduced.
The resulting gauge-invariant term goes as $(c^{(3)}_{l}/\Lambda^2) (\bar{L}^c\,  L) \Phi^{\star} \Phi\, \Delta$.
  When $\Delta$ obtains a vacuum expectation value $v_{\Delta}$, the above interaction represents an effective interaction between neutrinos and DM as in eq.~(\ref{op3}). 
Such an interaction can arise from the mediation of another scalar triplet with mass $\sim \Lambda$. 
The LEP constraint on the mass of the neutral scalar other than the SM-like Higgs, arising from such a Higgs triplet reads $m_{\Delta} \gtrsim 72$~GeV~\cite{Schael:2006cr}. Furthermore, 
theoretical bounds, constraints from $T$-parameter and Higgs signal strength in the diphoton channel dictate that $m_{\Delta} \gtrsim 150$~GeV~\cite{Das:2016bir} for $v_{\Delta} \sim 1$~GeV.
For smaller values of $v_{\Delta}$, such as $v_{\Delta} \sim 10^{-4}$~GeV,  the bound can be even stronger, $m_{\Delta} \gtrsim 330$~GeV. 
Moreover, the corresponding Wilson coefficient should be perturbative, $c^{(3)}_{l} \lesssim \sqrt{4 \pi}$. These two facts together lead to $c^{(3)}_{l}/\Lambda^2 \lesssim 2.5 \times 10^{-5}$~GeV$^{-2}$ for $\Lambda \sim m_{\Delta} \sim 150$~GeV. Such  values of $c^{(3)}_{l}/ \Lambda^2$ are rather small to lead to any significant flux suppression.
While this is true for a tree-level generation of  this interaction \textit{via} a triplet scalar exchange, such interactions can  be generated at the loop-level or by some new dynamics.

The renormalisable case corresponding to the effective interaction in eq.~(\ref{op3}) is discussed in greater detail in Sec.~\ref{crossS} and Sec.~\ref{CTR}.

\begin{figure}[h!]
    \centering
\includegraphics[width=2.8in,height=2.1in,angle=0]{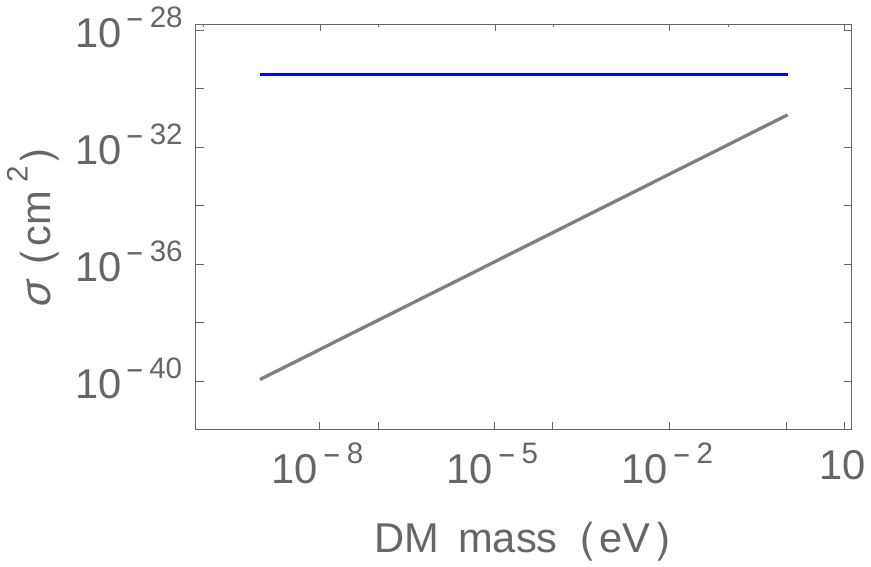}
 \caption{Cross-section vs. mass of DM. Blue line represents cross-section for $m_{\nu} = 0.1$~eV, $c^{(3)}_{l}/ \Lambda = 0.5$ GeV$^{-1}$ for interaction (3) under Topology~I. Grey line represents the required cross-section to induce $1\%$ suppression of incoming flux.}
    \label{fig:mesh}
\end{figure}

\item
There can also be a dimension-seven effective interaction vertex for neutrino-DM scattering: 
\bea
\mathcal{L} \supset  \frac{c_{l}^{(4)}}{\Lambda^3}(\bar{\nu^{c}} \sigma^{\m\n} \nu)(\partial_{\m} \Phi^{*} \partial_{\nu} \Phi - \partial_{\nu} \Phi^{*} \partial_{\mu} \Phi).
\label{O7}
\eea

Bound on this interaction comes from invisible $Z$ decay width and reads $c_{l}^{(4)}/\Lambda^3 \lesssim 2.0 \times 10^{-3}$~GeV$^{-3}$.
There is no counterpart of such an interaction involving the charged leptons. Thus the gauge-invariant form of this vertex does not invite any tighter bounds.
Such a bound dictates that this interaction does not lead to any  considerable flux suppression.

\item
 Another seven-dimensional interaction can be given as:
\bea
\mathcal{L} \supset  \frac{c_{l}^{(5)}}{\Lambda^3} \partial^{\m}(\bar{\nu^{c}}  \nu) \partial_{\m} (\Phi^{*}  \Phi).
\label{O8}
\eea

 From invisible $Z$ decay width the constraint on the coupling reads $c_{l}^{(5)}/\Lambda^3 \lesssim 7.5 \times 10^{-4}$~GeV$^{-3}$.
The measurement of $Z \ra l^{+}l^{-}$ or LEP monophoton searches does not invite any further constraint on this interaction due to the same reasons as in case of eq.~(\ref{op3}) and~(\ref{O7}).
Due to such a constraint, no significant flux suppression can take place in presence of this interaction.

\item
Another neutrino-DM interaction of dim-8 can be written as follows: 

\bea
\mathcal{L} \supset  \frac{c_{l}^{(6)}}{\Lambda^4}(\bar{\nu}  \partial^{\m} \gamma^{\nu} \nu)(\partial_{\m} \Phi^{*} \partial_{\nu} \Phi - \partial_{\nu} \Phi^{*} \partial_{\mu} \Phi).
\label{O6}
\eea

The coupling $c_{l}^{(6)}/\Lambda^4$ of interaction given by eq.~(\ref{O6}) is constrained from invisible $Z$ decay width as $c_{l}^{(6)}/\Lambda^4 \lesssim 2.5 \times 10^{-5}$~GeV$^{-4}$.
The constraint on the gauge-invariant form of this interaction reads $c_{l}^{(6)}/\Lambda^4 \lesssim  10^{-5}$~GeV$^{-4}$, which is similar for all three charged leptons. 
The gauge-invariant form of the above effective interaction also gives rise to five-point vertices involving the $Z$ boson. These lead to bounds from the observations of $Z \rightarrow inv$ and $Z \rightarrow l^{+}l^{-}$  which read $c_{l}^{(6)}/\Lambda^4 \lesssim 4.0 \times 10^{-5}$~GeV$^{-4}$ and $c_{l}^{(6)}/\Lambda^4 \lesssim 2.8 \times 10^{-5}$~GeV$^{-4}$ respectively.
The bound from the process $e^{+} e^{-} \rightarrow  \gamma \Phi^{*} \Phi$ reads $c_{e}^{(6)}/\Lambda^4 \lesssim 1.2 \times 10^{-6}$~GeV$^{-4}$.
Even with the least stringent constraint among the different considerations stated above, such an interaction does not lead to any significant flux suppression of the astrophysical neutrinos.

\end{enumerate} 

\subsection{Topology II} 

\label{Topb}

\begin{enumerate}
\item 
We consider a vector mediator $Z^{\prime}$, with couplings to neutrinos and DM given by:
\bea
\mathcal{L} \supset  \frac{c^{(7)}_{l}}{\Lambda^{2}}(\partial^{\mu}\Phi^{*} \partial^{\nu}\Phi - \partial^{\nu}\Phi^{*} \partial^{\mu}\Phi )Z'_{\mu \nu}+ f_i \bar{\n}_i \gamma^{\mu} P_L \n_{i} Z'_{\mu}.
\label{topb1}
\eea

This interaction has the same form of interaction as in eq~(\ref{O6}) of Topology I. Bound on this interaction from invisible $Z$ decay reads $f_l c_{l}^{(7)}/\Lambda^2 \lesssim 4.2 \times 10^{-2}$~GeV$^{-2}$.
The constraints on the gauge-invariant form of such interactions are $f_e c_{e}^{(7)}/\Lambda^2 \lesssim  5.8 \times 10^{-3}$~GeV$^{-2}$, $f_{\mu} c_{\mu}^{(7)}/\Lambda^2 \sim f_{\tau} c_{\tau}^{(7)}/\Lambda^2  \lesssim  8.1 \times 10^{-3}$~GeV$^{-2}$. The bound on the process $e^{+} e^{-} \rightarrow  \gamma \Phi^{*} \Phi$ reads $f_e c_{e}^{(7)}/\Lambda^2 \lesssim  1.9 \times 10^{-5}$~GeV$^{-2}$.

For this interaction, the $\Phi \Phi^{*} Z'$ vertex from eq.~(\ref{topb1}) takes the form, 
\bea
i \frac{c^{(7)}_{l}}{\Lambda^2}(p_2.p_4-m_{\text{DM}}^2)(p_2+p_4)^\mu Z'_\mu \, \sim \, i \frac{c^{(7)}_{l}}{\Lambda^2}m_{\text{DM}} (E_4-m_{\text{DM}})(p_2+p_4)^\mu Z'_\mu,\nn 
\eea 
where $p_{2}$ and $p_4$ are the four-momenta of the incoming and outgoing DM respectively.
In light of the constraints from $Z$ decay, the factor $\Big( c_{l}^{(7)} m_{\text{DM}}(E_4-m_{\text{DM}})/\Lambda^2\Big)$ is much smaller than unity when the dark matter is ultralight,\ie $m_{\text{DM}} \lesssim 1$ eV and incoming neutrino energy $\sim 1$ PeV. 
The rest of the Lagrangian is identical to the renomalisable vector-mediated process discussed in Sec.~\ref{crossV} and Sec.~\ref{CTR}. Further  
the charged counterpart of the second term in eq.~(\ref{topb1}) contributes to $g - 2$ of charged leptons and also leads to new three-body decay channels of $\tau$.
As mentioned in Sec.~\ref{vector}, the bounds on the these couplings read $f_e \sim 10^{-5}$, $f_{\mu} \sim 10^{-6}$ and $f_{\tau} \sim 10^{-2}$ for $m_{Z'} \sim 10$~MeV.  
So among the constraints from different considerations, even the least stringent one ensures that no significant flux suppression takes place with this interaction in case of ultralight DM.

\item
Consider a scalar mediator $\Delta$ with a momentum-dependent coupling with DM, 
\bea
\mathcal{L} \supset  \frac{c^{(8)}_{l}}{\Lambda} \partial^\mu |\Phi|^2 \partial_\mu \Delta +f_{l} \bar{\nu^{c}}  \nu \Delta.
\label{topb3}
\eea

Here $\Delta$ can be realised as the neutral component of a $SU(2)_L$-triplet scalar with $Y=2$. 
A Majorana neutrino mass term with $m_{\nu} = f_l v_{\Delta}$ also exists along with the second term of eq.~(\ref{topb3}), where $v_{\Delta}$ is the vev of the neutral component of the triplet scalar.
 The measurement of the $T$-parameter dictates, $v_{\Delta} \lesssim 4$~GeV~\cite{Agashe:2014kda}. For $v_{\Delta} \sim 1$~GeV, the smallness of neutrino mass constrains the coupling $f_{l}$ at $\sim \mathcal{O}(10^{-11})$. 
The mass of the physical scalar $\Delta$ is constrained to be $m_{\Delta}
\gtrsim 150~$GeV~\cite{Das:2016bir} for $v_{\Delta} \sim 1~$GeV.
For $f_{l} \sim \mathcal{O}(10^{-11})$ and $m_{\Delta}
\gtrsim 150~$GeV, such an interaction does not give rise to  an appreciable flux suppression for ultralight DM.

\end{enumerate}

\subsection{Topology III} 
\label{Topc}
We consider the vector boson $Z^{\prime}$ mediating the neutrino-DM interaction, with a renormalisable vectorlike coupling with the DM, but a non-renormalisable dipole-type interaction in the $\nu \nu Z^{\prime}$ vertex. The interaction terms are given as,
\bea
\mathcal{L} \supset C_{1}(\Phi^* \partial_\mu \Phi- \Phi \partial_\mu \Phi^*)Z'^\mu +  \frac{c^{(9)}_{l} }{\Lambda} (\bar{\nu^{c}} \sigma_{\mu \nu} P_L \n)Z'^{\mu \nu}.
\label{Of}
\eea
This interaction can be constrained from the measurement of the invisible decay width of $Z$. The flavour-independent bound on the coefficient $c_{l}^{(9)}$ reads, $c_{l}^{(9)}/\Lambda \lesssim 3.8\times 10^{-3}$~GeV$^{-1}$. The interactions in eq.~(\ref{Of}) can be realised as the renormalisable description of the effective Lagrangian as mentioned in eq.~(\ref{O7}).

\begin{figure}[h!]
    \centering
\subfigure[]{
\includegraphics[width=2.8in,height=2.1in,angle=0]{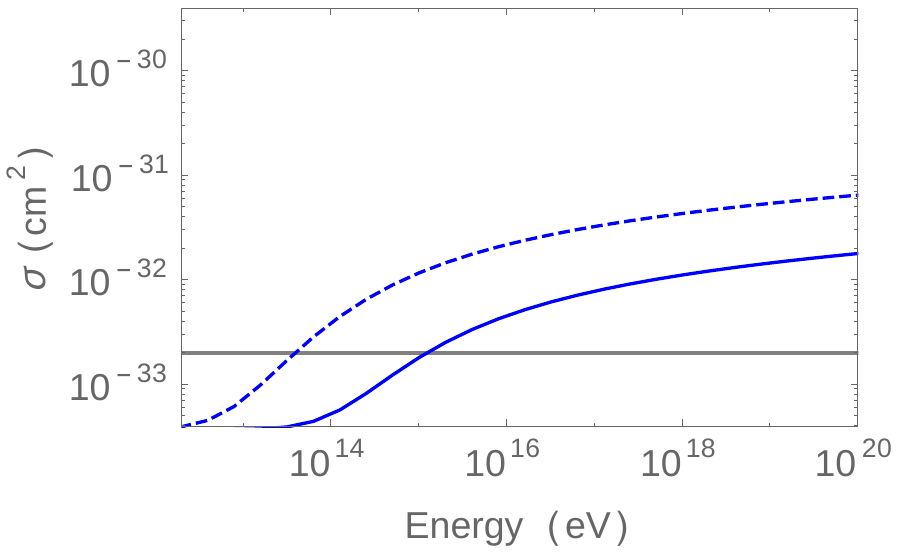}}
 \hskip 10pt
\subfigure[]{
\includegraphics[width=2.6in,height=2.1in,angle=0]{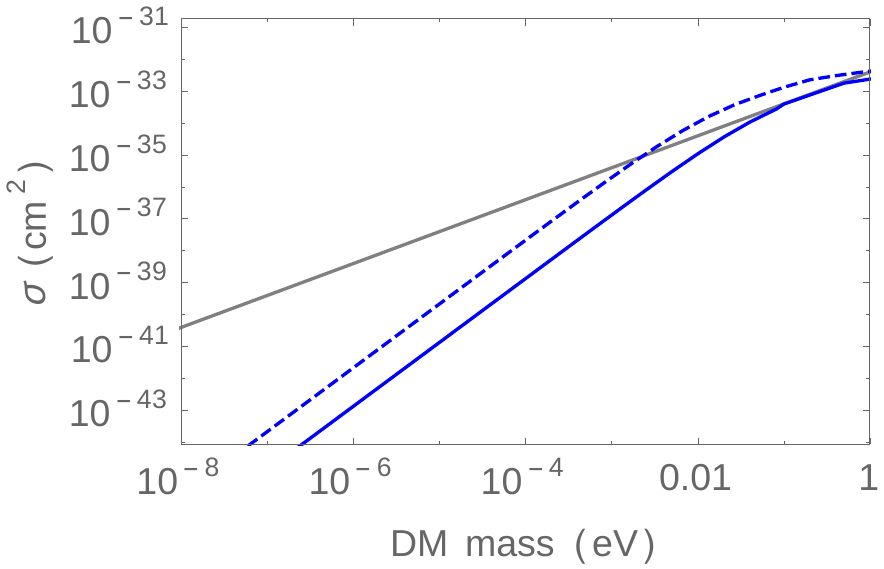}}
\caption{(a) Cross-section vs. incoming neutrino energy. (b) Cross-section vs. mass of DM. Grey line represents the required cross-section to induce $1\%$ suppression of incoming flux. The dashed and solid blue lines represent cross-sections for $m_{Z'} = 5,10$~MeV respectively, (a) with $m_{\text{DM}} =  0.5$~eV and (b) with $E_{\nu} =  1$~PeV. In both plots, $c_{l}^{(9)}/\Lambda =  3.8\times 10^{-3}$~GeV$^{-1}$ and $C_1 =  1$.}
    \label{fig:me}
\end{figure}

From fig.~\ref{fig:me} it can be seen that, for $m_{Z'} = 5,10~$MeV, such an interaction leads to a  significant flux suppression of neutrinos with energy $\sim 1~$PeV for DM mass in the range $0.002 - 1~$eV and $0.08 - 0.5$~eV respectively.

\subsection{Topology IV}
We consider the fermionic field $F_{L,R}$ mediating the neutrino-DM interaction with
\bea 
\mathcal{L} \supset \frac{c^{(10)}_{l}}{\Lambda^2}\bar{L} F_R \Phi |H|^2 + C_{L} \bar{L} F_R \Phi + h.c.
\label{topd}
\eea
In eq.~(\ref{topd}), after the Higgs $H$ acquires vacuum expectation value~(vev), the first term reduces to the second term up to a further suppression of ($v^2/\Lambda^2$). Following the discussion in Sec.~\ref{ren_ferm}, such interactions do not lead to a significant flux suppression.


\noindent $\bullet$ {\bf Effective interactions with thermal DM}\\
So far we have mentioned the constraints on several neutrino-DM interactions in case of ultralight DM and whether such interactions can lead to any significant flux suppression. 
Here we discuss such effective interactions of neutrinos with thermal DM with mass $m_{\text{DM}} \gtrsim 10$~MeV.
In case of thermal DM, bounds on the effective interactions considered above can come from the measurement of the relic density of DM, collisional damping and the measurement of the effective number of neutrinos, discussed in detail in Sec.~\ref{CTR}.
As mentioned earlier, the case of thermal DM becomes interesting in cases where the cross-section of neutrino-DM scattering increase with DM mass. For example, in topology~II with the interaction given by eq.~(\ref{topb1}), the neutrino-DM scattering cross-section is proportional to $\Big(c_{l}^{(7)} m_{\text{DM}}(E_4-m_{\text{DM}})/\Lambda^2\Big)$ which increases with DM mass. However, considering the bound on $c_{l}^{(7)}/\Lambda^2$ from $Z$ decay, the relic density and thus the number density of the DM with such an interaction comes out to be quite small, leading to no significant flux suppression. The following argument holds for all effective interactions considered in this paper for neutrino interactions with thermal DM. 
The thermally-averaged DM annihilation cross-section is given by $\langle\sigma v \rangle_{th} \propto (1/\Lambda^2)(m_{\text{DM}}^2 /\Lambda^2)^{d}$, where $d=0,1,2,3$ for five-, six-, seven- and eight-dimensional effective interactions respectively. 
In order to have sufficient number density, the DM should account for the entire relic density,\ie $\langle\sigma v \rangle_{th}  \sim 3\times10^{-26} \text{cm}^{3} \text{s}^{-1} $. 
To comply with the measured relic density, the required values of $\Lambda$ come out to be rather large leading to small cross section.

\section{The Renormalisable Models}
\label{renmodels}

\subsection{Description of the models}
Here we have considered three cases of neutrinos interacting with scalar dark matter at the tree-level \textit{via} a fermion, a vector, and a scalar mediator. 

\subsubsection{Fermion-mediated process} 
\label{ren_ferm}
In this case, the Lagrangian which governs the interaction between neutrinos and DM is given by:
\bea
\mathcal{L} \supset (C_L \bar{L} F_R  + C_R \bar{l}_R F_L)\Phi + h.c.
\label{fermren}
\eea
Here $L$ and $l_R$ stand for SM lepton doublet and singlet respectively. $F_{{L,R}}$ are the mediator fermions. 
As it was discussed earlier, a scalar DM of ultralight nature can only transform as a singlet under the SM gauge group. So, the new fermions $F_{L}$ and $F_R$ should transform as singlets and doublets under $SU(2)_L$ respectively. In such cases, the LEP search for exotic fermions with electroweak coupling lead to the bound on the masses of these fermions as, $m_F \gtrsim 100$~GeV~\cite{Achard:2001qw}. 
A scalar DM candidate can be both self-conjugate and non-self-conjugate. 
The stability of such DM can be ensured by imposing a discrete symmetry, for example, a $Z_2$-symmetry. 
A non-self-conjugate DM can be stabilised by imposing a continuous symmetry as well.
For self-conjugate DM, the neutrino-DM interaction takes place \textit{via} $s$- and $u$-channel processes and such contributions tend to cancel each other in the limit $s, u \ll m_{F}^2$~\cite{Boehm:2003hm}. 
In contrary, for non-self-conjugate DM the process is mediated only \textit{via} the $u$-channel and leads to a larger cross-section compared to the former case. 
In this paper, we only concentrate on the non-self-conjugate DM in this scenario.

Such interactions contribute to the anomalous magnetic moment, $\delta a_l \equiv g_{l} - 2$, of the charged SM leptons, which in turn constrains the value of the coefficients $C_{L,R}$.
The contribution of the interaction in eq.~(\ref{fermren}) to the anomalous dipole moment of SM charged lepton of flavour $l$ is given by~\cite{Leveille:1977rc}:   
\bea 
\delta a_l = \frac{m_{l}^{2}}{32\pi^{2}} \int_{0}^{1} dx \frac{(C_{L}+C_{R})^{2} (x^{2}-x^{3}+x^{2} \frac{m_{F}}{m_{l}})+ (C_{L}-C_{R})^{2} (x^{2}-x^{3}-x^{2} \frac{m_{F}}{m_{l}})}{m_{l}^{2} x^{2}+(m_{F}^{2}-m_{l}^{2})x+m_{\text{DM}}^{2}(1-x)},  
\eea
where $m_l$ is the mass of the corresponding charged lepton.
In the limit $m_{\text{DM}} \ll m_{l} \ll m_{F}$, the anomalous contribution due to new interaction reduces to,
\bea 
\delta a_l \sim \frac{C_LC_R m_{l}}{16 \pi^2 m_{F}}. 
\eea
For electron and muon the bound on the ratio $(C_L C_R / 16 \pi^2 m_{F})$ reads $1.6 \times 10^{-9}$~GeV$^{-1}$ and $2.9 \times 10^{-8}$~GeV$^{-1}$ respectively.
There is no such bound for the tauon.

\begin{figure}[h!]
 \begin{center}
\subfigure[]{
 \includegraphics[width=1.0in,height=1.5in, angle=0]{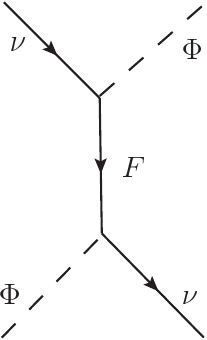}}
 \hskip 25pt
 \subfigure[]{
 \includegraphics[width=1.0in,height=1.5in, angle=0]{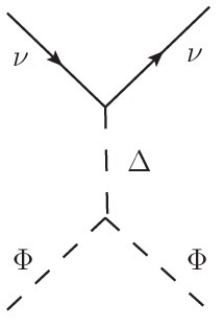}}
  \hskip 25pt
 \subfigure[]{
 \includegraphics[width=0.95in,height=1.5in, angle=0]{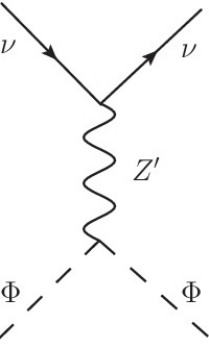}}

 \caption{Renormalisable cases of neutrino-DM scattering with (a) fermion, (b) scalar and (c) vector mediator. }
 \label{fynmnd}
\end{center} 
 \end{figure}

\subsubsection{Scalar-mediated process}
\label{Scalar}
The Lagrangian for the scalar-mediated neutrino-DM interaction can be written as:
\bea
\mathcal{L} \supset f_l \bar{L}^{c} L \Delta   + g_\Delta \Phi^* \Phi |\Delta|^2,
\label{eq}
\eea
where $L$ are the SM lepton doublets and $\Delta$ is the $SU(2)_{L}$-triplet with hypercharge $Y=2$.
When $\Delta$ acquires a vev $v_{\Delta}$, the first term in eq.~(\ref{eq}) leads to a non-zero neutrino mass $m_{\nu} \sim  f_{l} v_{\Delta}$.
For $v_{\Delta} \sim 1$~GeV and mass of the neutrino $m_{\n} \lesssim 0.1$~eV the  constraint on the coupling $f_l$ reads $f_{l} \lsim 10^{-11}$.
The second term in eq.~(\ref{eq}) contributes to DM mass $m_{\text{DM}}^2 \sim g_{\Delta} v_{\Delta}^{2}$. 
In case the DM mass is solely generated from such a term, the upper bound on $v_{\Delta}$ dictated by the measurement of $\rho$-parameter, implies a lower bound on $g_{\Delta}$.
The mass term for DM might also arise from some other mechanisms, for example, by vacuum misalignment in case of ultralight DM.
In such a scenario, for a particular value of $m_{\text{DM}}$ and $v_{\Delta}$ there exists an upper bound on the value of $g_{\Delta}$.

The lower bound on the mass of the heavy CP-even neutral scalar arising from the $SU(2)$-triplet is $m_{\Delta} \sim 150$~GeV for $v_{\Delta} \sim 1$~GeV~\cite{Das:2016bir}, which comes from the theoretical criteria such as perturbativity, stability and unitarity, as well as the measurement of the $\rho$-parameter and $h \rightarrow \gamma \gamma$. 

\subsubsection{Light $Z'$-mediated process} 
\label{vector}
The interaction of a scalar DM with a new gauge boson $Z'$ is given by the Lagrangian,
\bea
\mathcal{L} \supset f'_l \bar{L} \gamma^{\mu} P_L L Z'_{\mu} + ig^{\prime}(\Phi^{*} \partial^{\mu} \Phi-\Phi \partial^{\mu} \Phi^{*}) Z'_{\mu}. 
\label{vecren}
\eea
Here, $f'_l$ are the couplings of the $l=e, \mu, \tau$ kind of neutrinos with the new boson $Z^{\prime}$, while $g^{\prime}$ is the coupling between the dark matter and the mediator. $f'_l$ can be constrained from the $g - 2$ measurements. Due to the same reason as in the fermion-mediated case, the coupling of $Z^{\prime}$ with $\tau$-flavoured neutrinos is not constrained from $g-2$ measurements. Constraints for this case from the decay width of $Z$ boson will be discussed in Sec~V.

For the mass of the SM charged lepton, $m_{l}$ and the boson, $m_{Z'}$, the anomalous contribution to the $g-2$ takes the form~\cite{Leveille:1977rc}:
\bea 
\delta a_l \sim \frac{f_{l}^{\prime 2} m_{l}^2}{12 \pi^2 m_{Z'}^2}. 
\eea
We have considered vector-like coupling between the $Z^{\prime}$ and charged leptons. For electrons and muons we find the constraints on couplings-to-mediator mass ratio to be rather strong~\cite{Agashe:2014kda},
\bea
\frac{f'_e}{m_{Z'}} \lesssim \frac{7\times10^{-6}}{\text{MeV}}, \hspace{7pt}
\frac{f'_\mu}{m_{Z'}} \lesssim \frac{3\times10^{-7}}{\text{MeV}}.
\eea
From the measurement of $N_{\text{eff}}$ the lower bound on the mass of a light $Z'$ interacting with SM neutrinos at the time of nucleosynthesis reads $m_{Z'} \gtrsim 5~$MeV~\cite{Huang:2017egl}.

\subsection{Thermal Relic Dark Matter} 
\label{CTR}
In this scenario, the DM is initially in thermal equilibrium with other SM particles \textit{via} its interactions with the neutrinos. For models with thermal dark matter interacting with neutrinos, three key constraints come from the measurement of the relic density of DM, collisional damping and the measurement of the effective number of neutrinos. These three constraints are briefly discussed below. 

\vspace{10pt}

\noindent $\bullet$ {\bf Relic density} 

\noindent 
If the DM is thermal in nature, its relic density is set by the chemical freeze-out of this particle from the rest of the primordial plasma.
The observed value of DM relic density is $\Omega_{\text{DM}}h^{2} \sim 0.1188$~\cite{Agashe:2014kda}, which corresponds to the annihilation cross-section of the DM into neutrinos  $\bra\sigma v\ket_{th} \sim 3\times10^{-26} \text{cm}^{3} \text{s}^{-1}.$
In order to ensure that the DM does not overclose the Universe, we impose 
\bea
\langle\sigma v \rangle_{th}  \gtrsim 3\times10^{-26} \text{cm}^{3} \text{s}^{-1}.
\eea


\noindent $\bullet$ {\bf Collisional damping}

\noindent  Neutrino-DM scattering can change the observed CMB as well as the structure formation. In presence of such interactions, neutrinos scatter off DM, thereby erasing small scale density perturbations, which in turn suppresses the matter power spectrum and disrupts large scale structure formation. 
The cross-section of such interactions are constrained by the CMB measurements from Planck and Lyman-$\a$ observations as~\cite{Wilkinson:2014ksa,Escudero:2015yka}, 
\bea
\sigma_{el} \lsim 10^{-48}\times \Big(\frac{m_{\text{DM}}}{\text{MeV}}\Big)\Big(\frac{T_{0}}{2.35\times10^{-4}\text{eV}}\Big)^{2} \text{cm}^{2}.
\eea 

\vspace{10pt}

\noindent $\bullet$ {\bf Effective number of neutrinos}

\noindent 
In standard cosmology, neutrinos are decoupled from the rest of the SM particles at a temperature $T_{dec} \sim 2.3$~MeV and the effective number of neutrinos is evaluated to be $N_{\text{eff}} = 3.045$~\cite{deSalas:2016ztq}. For thermal DM in equilibrium with the neutrinos even below $T_{dec}$, entropy transfer takes place from dark sector to the neutrinos, which leads to the bound $m_{\text{DM}} \gtrsim 10$~MeV from the measurement of $N_{\text{eff}}$.
It can be understood as follows. In presence of $n$ species with thermal equilibrium with neutrinos, the change in $N_{\text{eff}}$ is encoded as~\cite{Boehm:2013jpa},
\bea
N_{\rm eff} = \Big(\frac{4}{11} \Big)^{-4/3} \Big( \frac{T_{\n}}{T_{\gamma}}\Big)^4 \Big[N_{\n} + \sum_{i=1}^{n}I\Big(\frac{m_{i}}{T_{\n}}\Big)\Big],
\eea
where,
\bea
\frac{T_{\n}}{T_{\gamma}} = \Big[\Big(\frac{g^{*}_{\nu}}{g^{*}_{\gamma}}\Big)_{T_{dec}} \frac{g^{*}_{\gamma}}{g^{*}_{\n}}\Big]^{1/3}.
\eea
Here, the effective number of relativistic degrees of freedom in thermal equilibrium with neutrinos is given as 
\bea
g_{\nu}^{*} = \frac{14}{8}\Big(N_{\nu} + \sum_{i=1}^{n} \frac{g_i}{2} F\Big(\frac{m_{i}}{T_{\n}}\Big)\Big).
\eea
In eqs.~(4.10) and (4.12), $i=1,..,n$ denotes the number of species in thermal equilibrium with neutrinos, $g_i = 7/8~(1)$ for fermions~(bosons) and the functions $I(m_{i}/T_{\n})$ and $F(m_{i}/T_{\n})$ can be found in ref.~\cite{Boehm:2013jpa}.  For a DM in thermal equilibrium with neutrinos and $m_{\text{DM}} \lsim 10$~MeV, the contribution of $F(m_{\text{DM}}/T_{\n})$ to $(T_{\n}/T_{\gamma})$ is quite large, and such values of DM mass can be ruled out from $N_{\text{eff}} = 3.15 \pm 0.23$~\cite{Ade:2015xua}, obtained from the CMB measurements.

We implement the above constraints in cases of the renormalisable models discussed in Sec~\ref{renmodels}. We present the thermally-averaged annihilation cross-section $\bra\sigma v\ket_{th}$ and the cross-section for elastic neutrino-DM scattering $\sigma_{el}$ for the respective models in table~\ref{tab:table0}. The notations for the couplings and masses follow that of  Sec~\ref{renmodels}. In the expressions of $\bra\sigma v\ket_{th}$, $p_{cm}$ can be further simplified as $\sim m_{\text{DM}} v_{r}$ where $v_r \sim 10^{-3}~c$ is the virial velocity of DM in the galactic halo~\cite{Campo:2017nwh}. In the expressions of  $\sigma_{el}$, $E_{\nu}$ represents the energy of the incoming relic neutrinos which can be roughly taken as the CMB temperature of the present Universe.

Two of the three renormalisable interactions discussed in this paper, namely the cases of fermion and vector mediators have been discussed in the literature in light of the cosmological constraints,\ie relic density, collisional damping and $N_{\text{eff}}$~\cite{Campo:2017nwh}.
For a particular DM mass, the annihilation cross-section decreases with increasing mediator mass. Thus, in order for the DM to not overclose the Universe, there exists an upper bound to the mediator mass for a particular value of $m_{\text{DM}}$. 
With mediator mass less than such an upper bound, the relic density of the DM is smaller compared to the observed relic density, leading to a smaller number density.

\begin{table}[h!]
	\centering
	\begin{tabular}{|c|c|c|c|}
        \hline
		 & Fermion-mediated & Scalar-mediated & Vector-mediated   \\
		\hline
		$\bra\sigma v\ket_{th}$     &   $C_{L}^{4}\frac{p_{cm}^{2}}{12 \pi(m_{\text{DM}}^{2}+m_{F}^{2})^{2}}$      & $g_{\Delta}^{2}f_{l}^{2}\frac{2m_{\text{DM}}^{2}+p_{cm}^{2}}{32 \pi m_{\text{DM}}^{2}(m_{\Delta}^{2}-4m_{\text{DM}}^{2})^{2}}$ & $g'^{2}f'^{2}\frac{p_{cm}^{2}}{3 \pi(m_{Z'}^{2}-4m_{\text{DM}}^{2})^{2}}$           \\
		
		$\sigma_{el} $     &   $C_{L}^{4}\frac{E_{\nu}^{2}}{8 \pi(m_{\text{DM}}^{2}-m_{F}^{2})^{2}}$       & $\frac{g_{\Delta}^{2}f_{l}^{2} E_{\nu}^{2}}{8 \pi m_{\text{DM}}^{2}  m_{\Delta}^{4}}$&$\frac{g'^{2}f'^{2} E_{\nu}^{2}}{2 \pi  m_{Z'}^{4}}$\\
		\hline
		
			\end{tabular}
\caption{Thermally averaged DM annihilation cross-section and the cross-section for neutrino-DM elastic scattering for thermal DM.}
	\label{tab:table0}
\end{table}

As  discussed earlier, the measurement of  $N_{\text{eff}}$ places a lower bound on DM mass $m_{\text{DM}} \gtrsim 10$~MeV. 
DM number density is proportional to the relic abundance and inversely proportional to the DM mass. 
Thus the most `optimistic' scenario in context of flux suppression is when $m_{\text{DM}} = 10~$MeV and the masses of the mediators  are chosen in such a way that those correspond to the entire relic density in fig.~5.
Such a choice leads to the maximum DM number density while satisfying the constraint of relic density and $N_{\text{eff}}$.
As it can be seen from fig.~5, such values of mediator and DM mass satisfies the constraint from collisional damping as well. 
 For example, as fig.~5(a) suggests, $m_{\text{DM}}\sim 10$~MeV and $m_{F} \sim 2$~GeV correspond to the upper boundary of the blue region, which represents the point of highest relic abundance. 
 Similarly for the scalar and vector mediated case, the values of mediator masses come out to be $\sim 20$~MeV and $\sim 1$~GeV respectively for  $m_{\text{DM}}\sim 10$~MeV.

 With the above-mentioned values of the DM and mediator masses, the neutrino-DM scattering cross-section for the entire range of energy of astrophysical neutrinos fall short of the cross-section required to produce $1\%$ flux suppression, by many orders of magnitude.  
 The key reason behind this lies in the fact that for the range of allowed
DM mass, corresponding number density is quite small and the neutrino-DM scattering cross-section cannot compensate for that.
 The cross-section in the fermion and scalar mediated cases decrease with energy in the relevant energy range. Such a fall in cross-section is much more faster in the scalar case compared to the fermion one. Though in the vector-mediated case the cross-section remains almost constant in the entire energy range under consideration.
  The cross-section in the fermion, scalar and vector-mediated cases are respectively $10^6 - 10^8$, $10^{30} - 10^{35}$ and $10^7$ orders smaller than the required cross-section in the energy range of $20$~TeV - $10$~PeV.
  Thus we conclude that the three renormalisable interactions stated above do not lead to any significant flux suppression of astrophysical neutrinos in case of cold thermal dark matter. 

\begin{figure}[h!]
 \begin{center}
\subfigure[]{
 \includegraphics[width=50mm,scale=0.5]{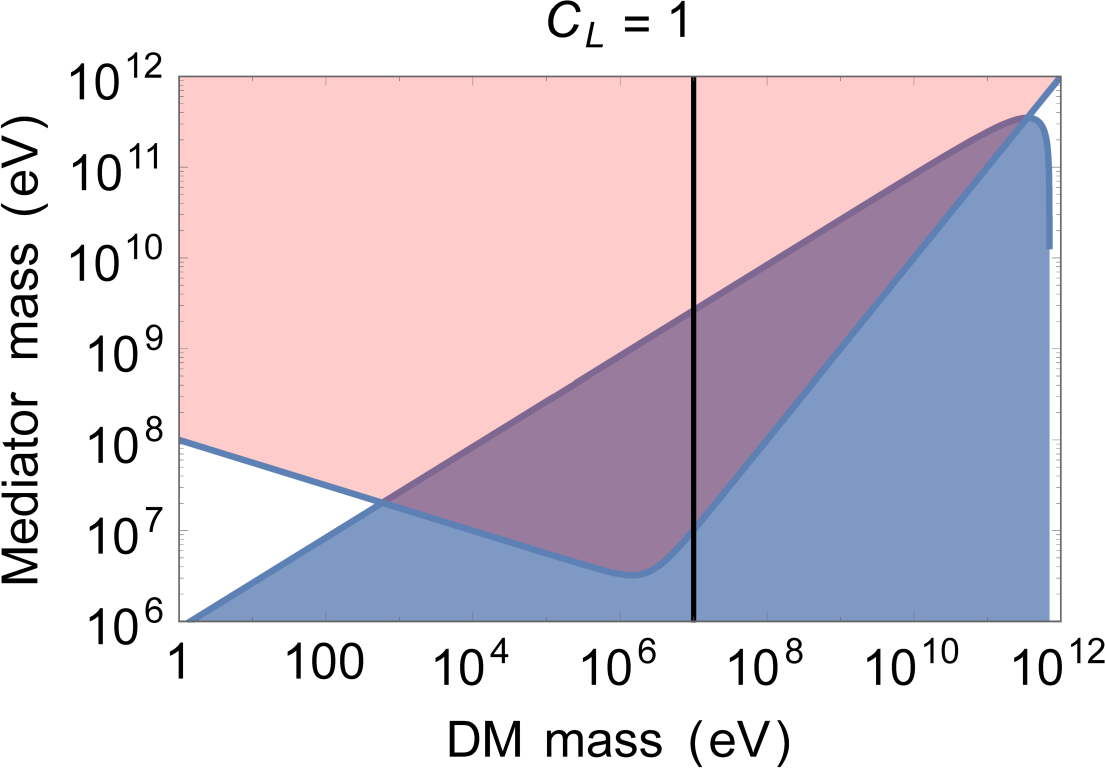}}
 \hskip 5pt
 \subfigure[]{
 \includegraphics[width=50mm,scale=0.5]{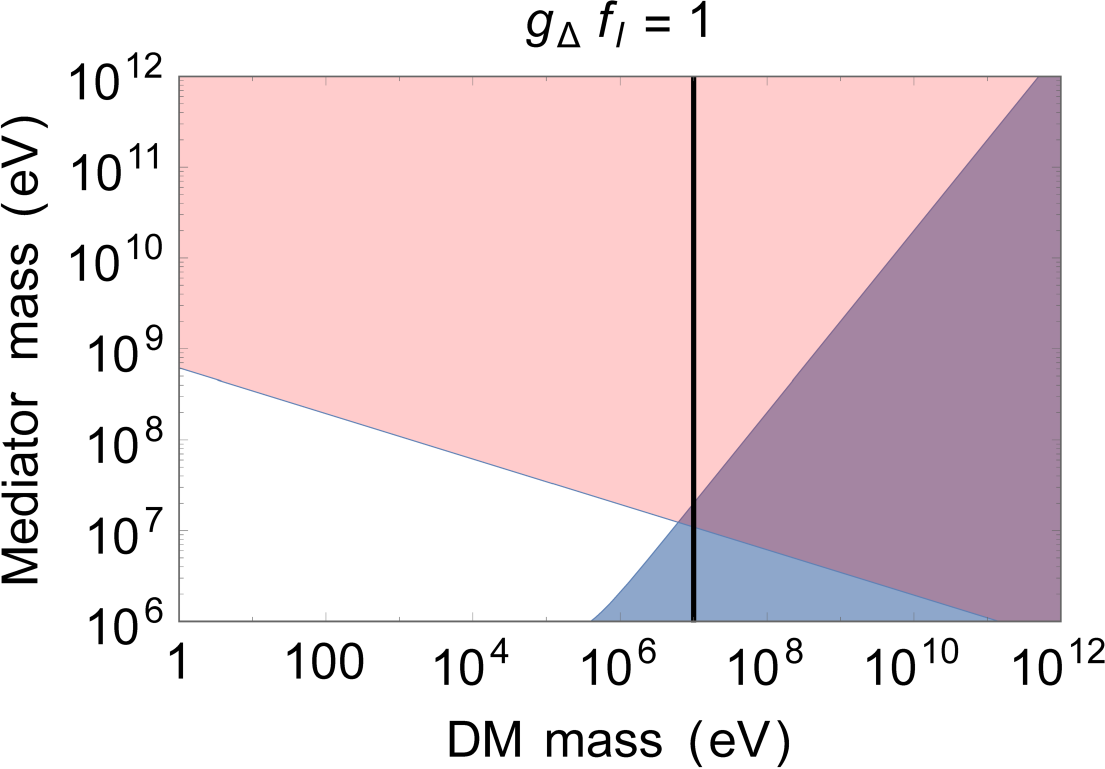}}
  \hskip 5pt
 \subfigure[]{
 \includegraphics[width=50mm,scale=0.5]{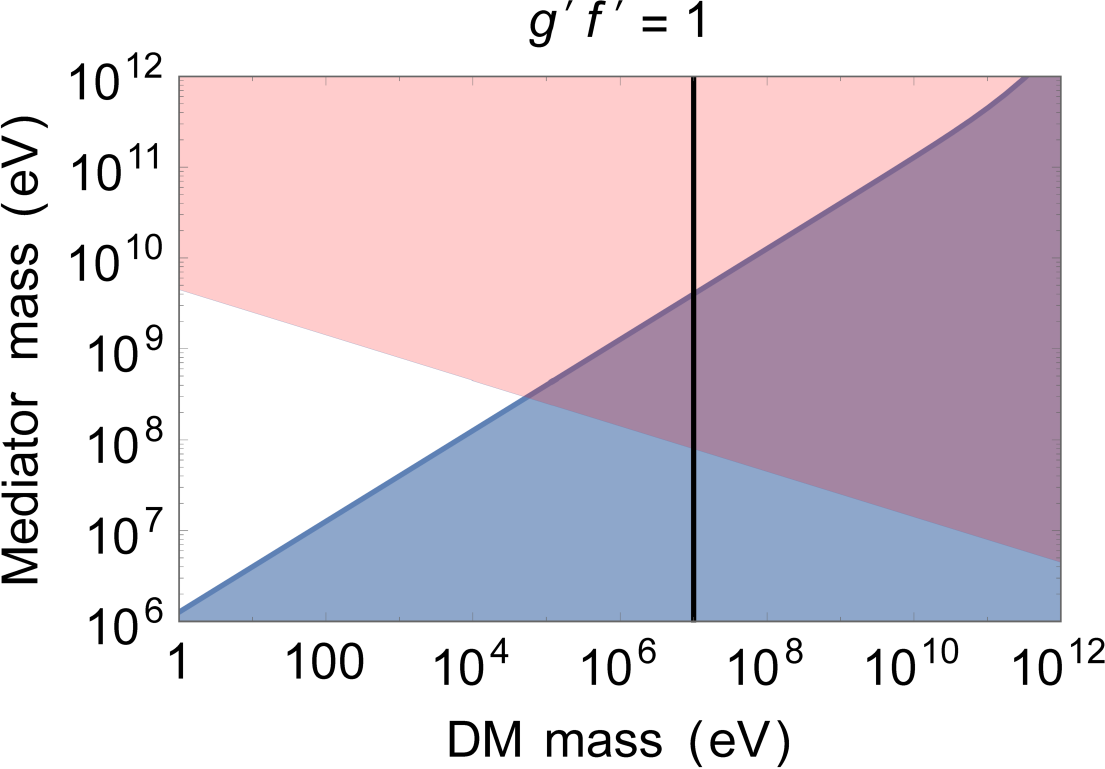}}

 \caption{Mass of the mediator vs. mass of DM for (a) fermion-mediated, (b) scalar-mediated and (c) vector-mediated neutrino-DM interactions. The blue and pink regions are allowed from relic density of DM  and collisional damping respectively. The region at the left side of the vertical black line is ruled out from the constraint coming from $N_{\rm eff}$.}
 \label{fig:mesh1}
\end{center} 
 \end{figure}

%

\subsection{Ultralight Scalar Dark Matter}
Here we consider the DM to be an ultralight BEC scalar with mass $10^{-21} - 1$~eV. 
The centre-of-mass energy for the neutrino-DM interaction in this case always lies between $\mathcal{O}(10^{-3})$~eV to $\mathcal{O}(10)$~MeV for incoming neutrino of energy $\sim \mathcal{O}(10)$~PeV  depending on DM mass. We consider below the  models described in Sec.~\ref{renmodels} to calculate the cross-section of neutrino-DM interaction and compare those to the cross-section required for a flux suppression at IceCube.
\subsubsection{Fermion-mediated process}
\label{crossF}

\begin{figure}[h!]
    \centering
\subfigure[]{
\includegraphics[width=2.8in,height=2.1in,angle=0]{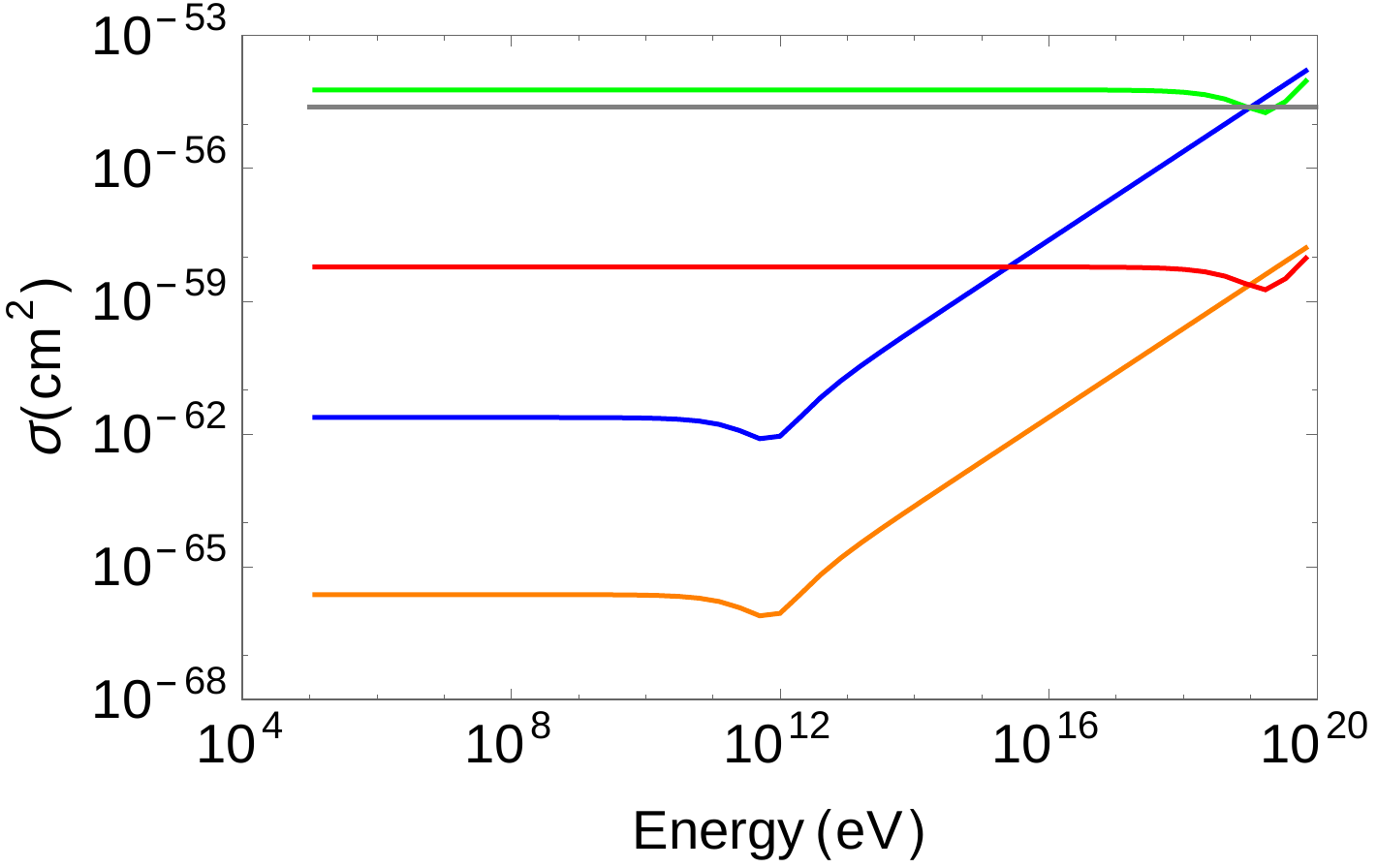}}
 \hskip 10pt
\subfigure[]{
\includegraphics[width=2.8in,height=2.1in,angle=0]{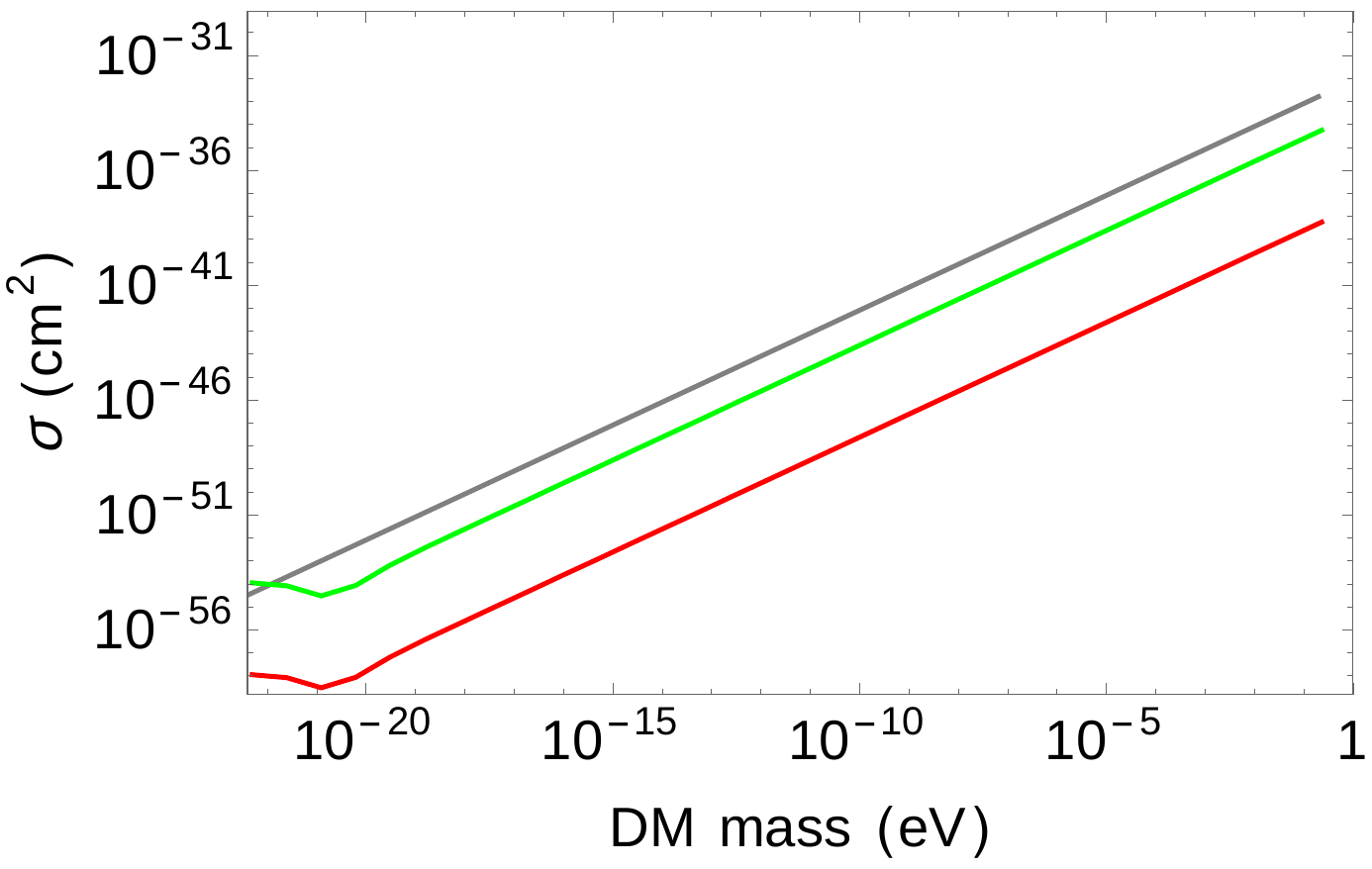}}
\caption{Fermion-mediated neutrino-DM scattering. (a) Cross-section vs. incoming neutrino energy. Green and blue lines represent cross-sections for  $m_{\n} =  10^{-2}, 10^{-5}$~eV respectively with $m_{F} =  10$~GeV. Red and orange lines represent cross-sections for $m_{\n} =  10^{-2}, 10^{-5}$~eV respectively with $m_{F} = 100$~GeV. Here $C_{L} = 0.88$, $m_{\text{DM}} = 10^{-22}$~eV. (b) Cross-section vs. mass of DM. Green and red lines represent $m_{F} = 10, 100$~GeV respectively for $m_{\nu} = 10^{-2}$~eV. Here $C_{L} = 0.88$, $E_{\n} = 1$~PeV. Grey line represents the required cross-section to induce $1\%$ suppression of incoming neutrino flux.}
    \label{fig:mesh3}
\end{figure}

The cross-section for neutrino-DM scattering through a fermionic mediator in case of ultralight scalar DM is given as 
\bea
\sigma \sim \frac{C_{L}^{4} (m_{\nu}^{2}+4 m_{\text{DM}} E_{\nu})}{16 \pi m_{F}^{4}}, \nn
\eea 
where $m_{\nu}, E_{\nu}$ are the mass and energy of the incoming neutrino respectively, $m_{\text{DM}}$ is the mass of the ultralight DM, and $m_F$ is the mass of the heavy fermionic mediator. 
As the mass of the DM is quite small, at lower neutrino energies $m_{\nu}^{2} > m_{\text{DM}} E_{\nu}$ and  hence, the cross-section remains constant. As the energy increases, the $m_{\text{DM}} E_{\nu}$ term becomes more dominant and eventually, the cross-section increases with energy.

Such an interaction has been studied in literature in case of ultralight DM~\cite{Barranco:2010xt}. This analysis was improved with the consideration of non-zero neutrino mass in ref.~\cite{Reynoso:2016hjr}. 
For example, from fig.~\ref{fig:mesh3}(a) it can be seen that the cross-section for  $m_{\n} \sim 10^{-2}$~eV is larger compared to that for $m_{\n} \sim 10^{-5}$~eV. In fig.~\ref{fig:mesh3}(b), with  $m_{\n} \sim 10^{-2}$~eV, it is shown that no significant flux suppression takes place for a DM heavier than $10^{-22}$~eV for $m_{F} \sim 10$~GeV.
However, it has been shown that the quantum pressure of the particles of mass $\lesssim 10^{-21}$ eV suppresses the density fluctuations relevant at small scales $\sim 0.1$~Mpc, which is disfavoured by the Lyman-$\a$ observations of the intergalactic medium~\cite{Irsic:2017yje,Armengaud:2017nkf}.
Also, the constraint on the mass of such a mediator fermion,  which couples to the $Z$ boson with a coupling of the order of electroweak coupling, reads $m_F \gtrsim 100$~GeV~\cite{Achard:2001qw}.
These facts together suggest that $m_{\text{DM}} \sim 10^{-22}$~eV and $m_F \sim 10$~GeV,  as considered in ref.~\cite{Reynoso:2016hjr}, are in tension with Lyman-$\alpha$ observations and LEP searches for exotic fermions respectively. 
If we consider $m_{\nu} = 0.1~$eV along with $m_{F} = 100~$GeV, it leads to a larger cross-section compared to that with $m_{\nu} = 0.01~$eV, which is still smaller compared to the cross-section required to induce a significant flux suppression.
Thus, taking into account such constraints, the interaction in eq.~(\ref{fermren}) does not lead to any appreciable flux suppression in case of ultralight DM.

\subsubsection{Scalar-mediated process}
\label{crossS}
\begin{figure}[h]
    \centering
\includegraphics[width=2.8in,height=2.1in,angle=0]{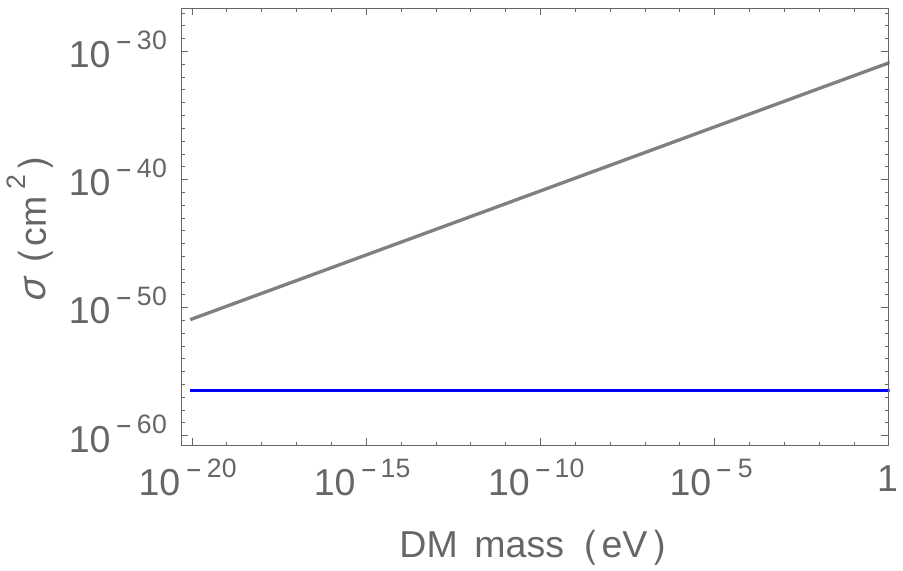}
\caption{Cross-section vs. mass of DM in scalar-mediated neutrino-DM scattering. The blue and grey lines represent the cross-section with scalar mediator and the same required to induce $1\%$ suppression of incoming flux respectively. Here, energy of incoming neutrino $E_\n = 1$~PeV, mediator mass $m_{\Delta} = 200$~GeV and $f_{l} g_{\Delta} v_{\Delta} = 0.1$~eV.}
    \label{fig:mesh4}
\end{figure}

As mentioned in Sec.~\ref{Scalar}, the bound on the coupling of a scalar mediator $\Delta$ with neutrinos is quite stringent, $f_{l} v_{\Delta} \lesssim 0.1$~eV.  
Moreover, the mass of such a mediator are constrained as $m_{\Delta} \gtrsim  150$~GeV~\cite{Das:2016bir}.
In this case, the cross-section of neutrino-DM scattering is independent of the DM as well as the  neutrino mass for neutrino energies under consideration. 
As fig.~\ref{fig:mesh4} suggests, the neutrino-DM cross-section in this case never reaches the value of cross-section required to induce a significant suppression of the astrophysical neutrino flux for $m_{\text{DM}} \gtrsim 10^{-21}$~eV. 
As mentioned earlier, DM of mass smaller than $\sim 10^{-21}$~eV are disfavoured from Lyman-$\a$ observations.

\subsubsection{Vector-mediated process}
\label{crossV}
\begin{figure}[h]
    \centering
\subfigure[]{
\includegraphics[width=2.8in,height=2.1in,angle=0]{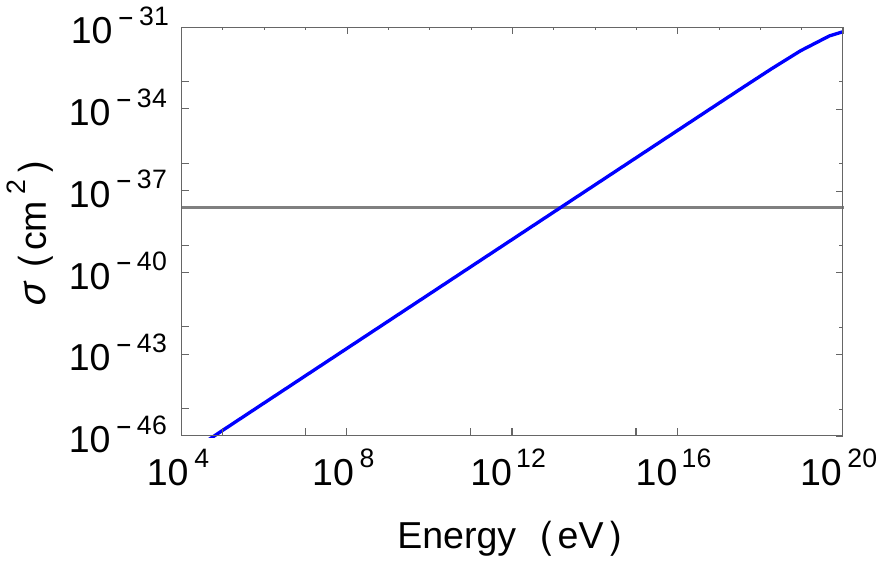}}
 \hskip 10pt
\subfigure[]{
\includegraphics[width=2.8in,height=2.1in,angle=0]{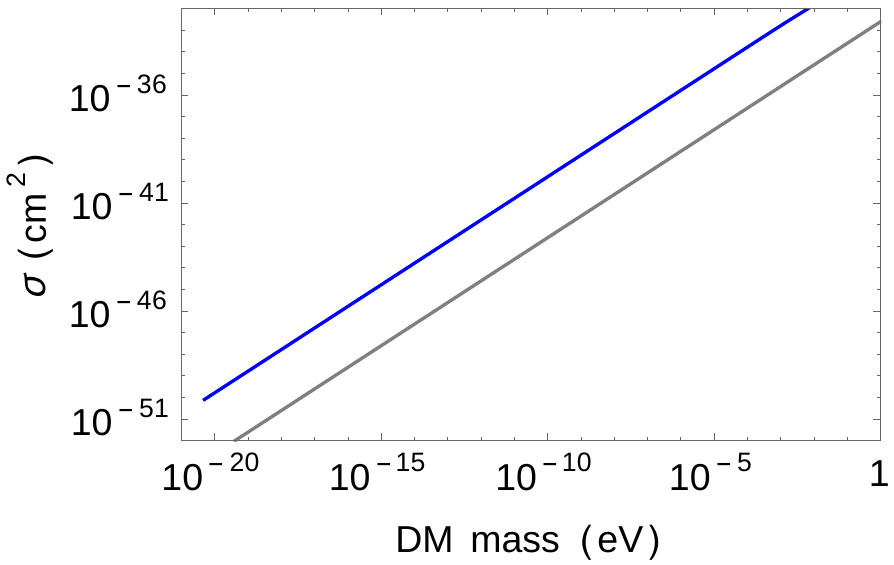}}
\caption{Vector-mediated neutrino-DM scattering. (a) Cross-section vs. incoming neutrino energy. (b) Cross-section vs. mass of DM. Blue and grey lines represent the calculated cross-section and required cross-section to induce $1\%$ suppression of incoming flux respectively. Here, the mediator mass $m_{Z^{\prime}} = 10$~MeV, and the couplings $g^{\prime} f^{\prime} = 10^{-3}$. For (a), $m_{\text{DM}} = 10^{-6}$~eV and for (b), the energy of incoming neutrino $ E_{\nu} = 1$~PeV.}
    \label{fig:mesh9}
\end{figure}

As it has been discussed in Sec.~\ref{vector}, the couplings of electron and muon-flavoured neutrinos to the $Z'$ are highly constrained, $\sim \mathcal{O}(10^{-5} - 10^{-6})$. However, as it will be discussed in Sec.~V, for the tau-neutrinos such a coupling is less constrained, $\sim \mathcal{O}(10^{-2})$. 
From fig.~\ref{fig:mesh9}(a)  it can be seen that, in presence of such an interaction, an appreciable flux suppression can take place for $E_{\nu} \gtrsim 10$~TeV, with $m_{Z'} = 10$~MeV, $g^{\prime} f^{\prime} = 10^{-3}$ and $m_{\text{DM}} = 10^{-6}$~eV.
Instead, if we fix $E_{\nu} = 1~$PeV, it can be seen from fig.~\ref{fig:mesh9}(b) that the entire range of DM mass in the ultralight regime,\ie $10^{-21}$~eV to $1$~eV, can lead to an appreciable flux suppression.
In the next section, we present a UV-complete model that can provide  such a coupling between the mediator and neutrinos in order to obtain a cross-section which leads to an appreciable flux suppression.

In the standard cosmology neutrinos thermally decouple from electrons, and  thus from photons, near $T_{dec} \sim 1$~MeV. 
Ultralight DM with mass $m_{\text{DM}}$ forms a Bose-Einstein condensate below a critical temperature $T_c = 4.8 \times 10^{-4}/\big((m_{\text{DM}}(\text{eV}))^{1/3}a \big)$~eV, where $a$ is the scale factor of the particular epoch~\cite{Das:2014agf}. When the temperature of the Universe is $T \sim T_{dec}$, $T_c \sim 480$~MeV for $m_{\text{DM}} \sim 10^{-6}$~eV,\ie the ultralight DM exists as a BEC.
In order to check whether the benchmark scenario  presented in fig.~\ref{fig:mesh9}(a) leads to late kinetic decoupling of neutrinos, we verify if $n_{\nu}(T_{dec}) \, \sigma_{\n-\text{DM}} \, v_{\n} \lsim H(T_{dec})$. Here, $n_{\nu}(T)$ and $H(T)$ are  the density of relic neutrinos and the Hubble rate at temperature $T$ respectively,
\bea
H(T_{dec}) &\sim & \frac{\pi\sqrt{g_{eff}}}{\sqrt{90}}\frac{T_{dec}^2}{M_{Pl}} 
 \sim 5 \times 10^{-16}~\text{eV}. \nn\\
 n_{\n} &\sim & 0.091 T_{dec}^3 \sim 1.14 \times 10^{31}~\text{cm}^{-3}.
\eea
For $m_{\text{DM}} \sim 10^{-6}$~eV, mediator mass $m_{Z'} \sim 10$~MeV and neutrino-DM coupling $g^{\prime} f^{\prime} \sim  10^{-3}$, $\sigma_{\nu-\text{DM}} \sim 1.5 \times 10^{-44}$~cm$^{-2}$.  Thus, at $T \sim T_{dec}$, $ n_{\n} \,  \sigma_{\nu-\text{DM}} \, v_{\nu} \sim 4.2 \times 10^{-20}$~eV $\ll H(T)$ with $v_{\nu}\sim c$.
This reflects that the neutrino-DM interaction in our benchmark scenario does not cause late kinetic decoupling of neutrinos. 
However, as fig.~\ref{fig:mesh9}(b) suggests, for a particular neutrino energy the neutrino-DM cross-section is sizable for higher values of $m_{\text{DM}}$, that can lead to late neutrino decoupling and we do not consider such values of $m_{\text{DM}}$.
It was also pointed out that a strong neutrino-DM interaction can degrade the energies of neutrinos emitted from core collapse Supernovae and scatter those off by an significant amount to not be seen at the detectors~\cite{Mangano:2006mp,Fayet:2006sa,Bertoni:2014mva}.
This imposes the following constraint on the neutrino-DM cross-section~\cite{Boehm:2013jpa,Mangano:2006mp}:  $\sigma_{\nu-\text{DM}} \lesssim 3.9 \times 10^{-25}$~cm$^{-2}~(m_{\text{DM}}/$MeV) for $E_{\nu} \sim T_{\text{SN}} \sim 30$~MeV. It can seen from fig.~\ref{fig:mesh9}(a) that such a constraint is comfortably satisfied in our benchmark scenario.

\section{A UV-complete model for vector-mediated ultralight scalar DM}
\label{zprime}
Here we present a UV-complete scenario which accommodates an ultralight scalar DM as well as a $Z^{\prime}$ with mass $\sim \mathcal{O}(10)$~MeV. The $Z^{\prime}$ mediates the interaction between the DM and neutrinos.

The coupling of such a $Z^{\prime}$ with the first two generations of neutrinos cannot be significant because of the stringent constraints on the couplings of the $Z^{\prime}$  with electron and the muon. As it was discussed  in Sec.~\ref{vector}, those couplings have to be $\sim \mathcal{O}(10^{-5} - 10^{-6})$ for $m_{Z^{\prime}} \sim 10$~MeV. Thus, only the couplings to the third generation of leptons can be sizable. However, the coupling of the $Z^{\prime}$ with the $b$-quark is also constrained from the invisible decay width of $\Upsilon$. 
The bound from such invisible decay width dictates $|g_{\Phi} g_{b}| \lesssim  5\times 10^{-3}$, where $g_{\Phi}$ and $g_{b}$ stand for $Z'$ coupling with DM and the $b$-quarks respectively~\cite{Fayet:2009tv}.
 Thus we construct a model such that the $Z^{\prime}$ couples only to the third generation of leptons among the SM particles. 

The $Z'$ boson is realised as the gauge boson corresponding to a $U(1)'$ gauge group, which gets broken at $\sim \mathcal{O}(10)$~MeV due to the vev of the real component of a complex scalar $\varphi$ transforming under the $U(1)'$.
 As the third generation of SM leptons are also charged under $U(1)^{\prime}$, in order to cancel the chiral anomalies it is necessary to include another generation of heavy chiral fermions to the spectrum~\cite{Frampton:1999xi}. 
The cancellation of chiral anomalies in presence of the fourth generation of chiral fermions under $SU(3)_c \times SU(2)_L \times U(1)_Y \times U(1)'$ is discussed in appendix~\ref{appendix:anomaly}.
If the exotic fermions obtain masses from the vev of the scalar $\varphi$ which is also responsible for the  mass of $Z'$, the mass of the exotic fermion is related to the gauge coupling of $U(1)'$ in the following manner~\cite{DiFranzo:2015qea,Dobrescu:2014fca},
 \bea
m_{\text{exotic}} \lesssim 100~\text{GeV} \Big(\frac{m_{Z'}}{10~\text{MeV}}\Big) \Big(\frac{5.4\times 10^{-4}}{g_{Z'}}\Big) \Big(\frac{1}{Y^{\prime}_\varphi}\Big).
\label{exo_pert} 
\eea
Here, $g_{Z'}$ is gauge coupling of $U(1)'$ and $Y^{\prime}_{\varphi}$ is the $U(1)'$ charge of the scalar $\varphi$. It is clear from eq.~(\ref{exo_pert}) that, in order to satisfy the collider search limit on the masses of exotic leptons $\sim 100$~GeV, the gauge coupling of  $Z'$ has to be rather small. 
Such a constraint can be avoided if the exotic fermions obtain masses from a scalar other than $\varphi$. 
This scalar cannot be realised as the SM  Higgs, because then the effect of the heavy fourth generation fermions do not decouple in the loop-mediated processes like $gg \rightarrow h$, $h \rightarrow \gamma\gamma$ etc. To evade both these constraints we consider that the exotic fermions get mass from a second Higgs doublet.

\begin{table}[h!]
	\centering
	\begin{tabular}{|c|c|c|c|c|c|}
	    \hline
		$\psi$ & $SU(3)_c$ & $SU(2)_L$ & $U(1)_Y$ & $U(1)'$ & $Z_2$ \\
		\hline
		$Q$     &   $3$       & $2$         & $1/6$   &   $0$  &   $+$    \\
		\hline
		$u_R$     &   $3$       & $1$         & $2/3$   &   $0$   &   $+$   \\
		\hline
		$d_R$     &   $3$       & $1$         & $-1/3$  &   $0$  &   $+$    \\
		\hline
		$L_{e},L_{\mu}$ &   $1$       & $2$         & $-1/2$  &   $0$  &   $+$    \\
		\hline
		$ e_{R}, \mu_{R} $&   $1$  & $1$         & $-1$   &   $0$   &   $+$   \\
		
		\hline
		$L_{\tau}$ &   $1$       & $2$         & $-1/2$ &   $1$   &   $+$   \\
		\hline
		$\tau_R$   &   $1$       & $1$         & $-1$  &   $1$   &   $+$   \\
		\hline
		
		$L_4$   &   $1$       & $2$         & $-1/2$ &  $-1$  &   $+$  \\
		\hline
		$l_{4R}$   &   $1$       & $1$         & $-1$   &   $-1$ &   $-$ \\
		\hline
	    $Q_{4}$   &   $3$       & $2$         & $1/6$ &  $0$  &   $+$  \\
		\hline
		$u_{4R}$   &   $3$       & $1$         & $2/3$   &   $0$ &   $-$ \\
		\hline
		$d_{4R}$   &   $3$       & $1$         & $-1/3$   &   $0$ &   $-$ \\
		\hline
		$\Phi_1$   &   $1$       & $2$         & $1/2$  &   $0$  &   $-$   \\
		\hline 
		$\Phi_{2}$ &   $1$       & $2$         & $1/2$  &   $0$  &   $+$ \\
		\hline
		$\varphi$  &  $1$       & $1$          & $0$    &   $Y_{\varphi}$  &   $+$  \\
		\hline
		$\nu_R$    &   $1$       & $1$         & $0$    &   $0$  &   $+$  \\ 
		\hline
        $\Phi$     &   1       & 1         & 0    &   $Y_\Phi$  &   $-$   \\ 
		\hline
	\end{tabular}
\caption{Quantum numbers of the particles in the model.}
	\label{content}
\end{table}

In order to avoid Higgs-mediated flavour-changing neutral current at the tree-level, it is necessary to ensure that no single type of fermion obtains mass from both the doublets $\Phi_{1,2}$. Hence, we impose a $Z_2$-symmetry to secure the above arrangement under which the fields transform as it is mentioned in table~\ref{content}.
After electroweak symmetry breaking, the spectrum of physical states of this model will contain two neutral CP-even scalars $h$ and $H$, a charged scalar $H^{\pm}$, and a pseudoscalar $A$. 
The Yukawa sector of this model looks like,
\bea
\mathcal{L_{\text{\tiny Yukawa}}} \supset \frac{m_{f}}{v}(\zeta^{f}_h \bar{f} f h+\zeta^{f}_H \bar{f} f H+\zeta^{f}_A \bar{f} f A),
\eea
with,
\bea
&&\zeta^{SM}_h = \cos \alpha/\sin \beta,\hspace{7pt}  \zeta^{SM}_H= \sin \alpha/\sin \beta,\hspace{7pt} 
\zeta^{SM}_A= -\cot \b, \nn\\
&&\zeta^{\chi}_h = -\sin \alpha/\cos \beta,\hspace{7pt} 
\zeta^{\chi}_H= \cos \alpha/\cos \beta,
\hspace{7pt}
\zeta^{\chi}_A= \tan \b.
\eea
Here, $\zeta^{SM}_i$ and $\zeta^{\chi}_i$ are the coupling multipliers of the SM and exotic fermions to the neutral scalars $i \equiv h,H,A$ respectively. 
It can be seen that the couplings of the Higgses with SM fermions in this model are the same as in a Type-I 2HDM.
$\a$ is the mixing angle between the neutral CP-even Higgses and $\b$ quantifies the ratio of the vevs of the two doublets, $\tan \b = v_2/v_1$.
The coupling of the SM-like Higgs to the exotic fermions tend to zero as $\a \rightarrow 0$. Moreover, the Higgs signal strength measurements dictate $|\cos (\b-\a)| \lesssim 0.45$ at $95\%$~CL~\cite{Dorsch:2016tab,Haber:2015pua}.
So, the allowed values of $\tan \b$ for our model are  $\tan \b \gtrsim 1.96$ along with $\a \rightarrow 0 $.
The particle content of our model along with their charges under the SM gauge group as well as $U(1)'$ and $Z_2$ are given in table~\ref{content}.
Chiral fourth generation fermions can also be realised in a Type-II 2HDM in  the wrong-sign Yukawa limit~\cite{Das:2017mnu}.

The $Z^{\prime}\tau\bar{\tau}$ interaction in our model leads to a new four-body decay channel of $\tau$ and three-body decay channels for $Z$ and $W^{\pm}$. We consider that the effect of these new interactions must be such that their contribution to the respective decay processes must be within the  errors of the measured decay widths at $1\sigma$ level.
This leads to an upper bound on the allowed value of the coupling $g_{\tau}$ which is enlisted in table~\ref{tab:table3}. 

\begin{table}[h!]
	\centering
	
	\begin{tabular}{|c|c|c|}
       \hline
		Process & Allowed decay width (GeV) & Maximum value of $g_{\tau}$  \\
		\hline
		$\tau \rightarrow \nu_{\tau} W^{-(*)} Z'$     &   $3.8 \times 10^{-15}$       & 0.04           \\
		\hline
		$W^{-} \rightarrow\tau^{-} \bar{\nu}_{\tau} Z' $     &   $1.8 \times 10^{-2}$       & 0.05\\
		\hline
		$Z \rightarrow \tau^{+} \tau^{-} Z'$     &   $2.8 \times 10^{-4}$       & $0.02$    \\
		\hline
			\end{tabular}
\caption{Constraints on coupling of light vector boson $Z'$ of mass $ 10$~MeV.}
	\label{tab:table3}
\end{table}

If we choose the new symmetry to be a $SU(2)$ instead of $U(1)'$, then in addition to $Z'$ we would have $ W'^{\pm}$ in the spectrum. 
But the existence of a charged vector boson of mass $\sim \mathcal{O}(10)$~MeV opens up a new two-body decay channel for $\tau$. Such decay processes are highly constrained, thus making the coupling of $Z'$ to $\n_{\tau}$ rather small.

\section{Summary and Conclusion}
\label{conclusion}
High energy extragalactic neutrinos travel a long distance before reaching Earth, through the isotropic dark matter background.
The observation of astrophysical neutrino flux at IceCube can bring new insights for a possible interaction between neutrinos and dark matter. 
While building models of neutrino-DM interactions leading to flux suppressions of astrophysical neutrinos, the key challenge is to obtain the correct number density of dark matter along with the required cross-section.
The number density of DM in the WIMP scenario is quite small compared to the ultralight case. 
However, the neutrino-DM scattering cross-section for some interactions increase with the DM mass.
Thus, it is essentially the interplay of DM mass and the nature of neutrino-DM interaction that collectively decide whether a model can lead to a significant flux suppression. 
So, a  study of various types of interactions for the whole range of DM masses is required to comment on which scenarios actually give the right combination of number density and cross-section.  

Issues of neutrino flux suppression~\cite{Barranco:2010xt, Reynoso:2016hjr,Arguelles:2017atb}, flavour conversion~\cite{deSalas:2016svi,Huang:2018cwo} and cosmological bounds~\cite{Boehm:2013jpa,Campo:2017nwh,Wilkinson:2014ksa,Escudero:2015yka} in presence of neutrino-DM interaction have been addressed in the literature.
The existing studies of the flux suppression of astrophysical neutrinos involve only a few types of renormalisable neutrino-DM interactions.
As mentioned earlier, such studies suffer from various collider searches and precision tests.
We take a rigourous approach to this problem by considering  renormalisable as well as effective interactions between neutrinos and DM and mention the constraints on such interactions.
Taking into account the bounds from precision tests, collider searches as well as the cosmological constraints, we investigate whether such interactions can provide the required value of cross-section of neutrino-DM scattering so that they lead to flux suppression of the astrophysical neutrinos.

In this paper we have contained our discussion to scalar dark matter.
Thermal DM with mass $m_{\text{DM}} \lesssim \mathcal{O}(10)$~MeV can be realised as warm and hot dark matter, whereas for $m_{\text{DM}} \gtrsim \mathcal{O}(10)$~MeV it can be realised as cold DM. 
However, non-thermal ultralight DM with mass in the range  $\mathcal{O}(10^{-21})$~eV -- $\mathcal{O}(1)$~eV can exist as a Bose-Einstein condensate,\ie as a cold DM as well. In contrary to the warm and hot thermal relics, which can only account for $\sim 1\%$ of the total DM density, ultralight BEC DM can account for the total DM abundance.
We consider three renormalisable interactions viz. the scalar, fermionic and vector mediation between neutrinos and DM at the tree-level.
Moreover, we consider up to dimension-eight contact type interactions in topology~I, and dimension-six interactions in one of the vertices in topologies~II, III and IV.
We find the constraints on such interactions from LEP monophoton searches, measurement of the $Z$ decay width and precision measurements such as anomalous magnetic dipole moment of $e$ and $\mu$.
In passing, we also point out that the demand of gauge invariance of the effective interactions can lead to more stringent constraints.  
For the thermal dark matter, we discuss the cosmological bounds on the models coming from relic density, collisional damping and measurement of effective number of neutrinos.

In case of thermal DM of mass greater than $\mathcal{O}(10)$~MeV, for a particular DM mass, the  value of mediator mass for renormalisable cases or the effective interaction scale for non-renormalisable cases, required to comply with the observed relic density, is too large to lead to a significant flux suppression of the astrophysical neutrinos. 
 For masses lower than $\mathcal{O}(10)$~MeV, the renormalisable neutrino-DM interaction \textit{via} a light $Z'$ mediator can lead to flux suppression of the high energy astrophysical neutrinos, that too only for $m_{\text{DM}} \lesssim 10$~eV.
For ultralight BEC DM, among effective interactions, one dim-5 contact-type interaction from topology~I and the dim-5 neutrino-dipole type interaction from topology~III give rise to  significant flux suppressions. Also, the renormalisable neutrino-DM interaction \textit{via} a light $Z'$ leads to an appreciable flux suppression.
 We present a UV-complete model accommodating the renormalisable neutrino-DM interaction in presence of such a light $Z'$ mediator.
We also discuss the need for a new generation of chiral fermions, a second Higgs doublet and a light scalar singlet to satisfy collider bounds and cancellation of chiral anomalies in such a consideration.  
Also we argue that, the benchmark scenario with a light $Z'$ mediator presented to demonstrate the flux suppression of high energy astrophysical neutrinos by ultralight DM, does not interfere with standard cosmological observables. 
The model presented at Sec.~V serves as an archetype of its kind, indicating the intricacies involved in such a model-building owing  to several competing constraints ranging from precision and Higgs observables to cosmological considerations.
A summary of all the interactions under consideration along with ensuing constraints, and remarks on relevance in context of high energy neutrino flux suppression can be found at appendix~\ref{summarytable}.

The effective neutrino-DM interactions considered in this paper can stem from different renormalisable models, at both tree and loop levels.
 In order to keep the analysis as general as possible, in contrary to the usual effective field theory~(EFT) prescription, we do not assume any particular scale of the dynamics which lead to such effective interactions. 
As a result, it is not possible to \textit{a priori} ensure that the effects of a particular neutrino-DM effective interaction will always be smaller than an effective interaction with a lower mass-dimension. 
Thus we investigate effective interactions up to mass dimension-8.

The possibility of neutrino-DM interaction in presence of light mediators, for example a $Z'$ with mass $\sim \mathcal{O}(10)$~MeV, points to the fact that effective interaction scale in such processes can be rather low.  
As it was mentioned earlier, the centre-of-mass energy for the scattering of the astrophysical neutrinos off ultralight DM particles can be quite small, $\sqrt{s} \lesssim 10$~MeV, for neutrino energies up to $1$~PeV. 
 Thus, it  might be tempting to try to interpret the effective interactions arising out of all the renormalisable scenarios with mediator mass $\gtrsim 10$~MeV,\ie even the case of a $Z'$ of mass $\sim \mathcal{O}(10)$~MeV, as higher-dimensional operators in an EFT framework.
 However, it has been shown that the $Z$-decay and LEP monophoton searches constrain both the renormalisable as well as effective neutrino-DM interactions.
Hence such an EFT description of neutrino-DM interaction does not hold below the $Z$ boson mass or $\sqrt{s}_{\text{LEP}} \sim 209$~GeV.
 For this reason, it is not meaningful to match the bounds   obtained in a renormalisable model with mediator mass less than $m_Z$ or $\sqrt{s}_{\text{LEP}}$,\ie the model with a light $Z'$ as in Sec.~\ref{crossV}, with the corresponding effective counterpart in eq.~(\ref{op2}).

It is also worth mentioning that the flavour oscillation length of the neutrinos is much smaller than the mean  interaction length with dark matter. 
Hence, the attenuation in the flux of one flavour of incoming neutrinos eventually gets transferred to all other flavours and leads to an overall flux suppression irrespective of the flavours.
The criteria of $1\%$ flux suppression helps to identify the neutrino-DM interactions which should be further taken into account to check potential signatures at IceCube. 
The flux of astrophysical neutrinos at IceCube also depend upon the specifics of the source flux and cosmic neutrino propagation. 
In order to find out the precise degree of flux suppression, one needs to solve an integro-differential equation consisting of both attenuation and regeneration effects~\cite{Naumov:1998sf}, which is beyond the scope of the present paper and is addressed in ref.~\cite{Karmakar:2018fno}.
But the application of the criteria of $1\%$ flux suppression, as well as the conclusions of the present work are independent of an assumption of a particular type of source flux or details of neutrino propagation.

In brief, we encompass a large canvas of interactions between neutrinos and dark matter, trying to find whether they can lead to flux suppression of the astrophysical neutrinos. 
The interplay of collider, precision and cosmological considerations affect such an endeavour in many different ways. 
Highlighting this, we point out the neutrino-DM interactions which can be probed at IceCube.

\vspace{20pt}

\noindent {\bf Note added}

\noindent The vector-mediated renormalisable interaction described in Sec.~\ref{crossV} can be subjected to strong BBN constraints: For $m_{Z'} \sim 10$~MeV, in order to avoid $s$-channel annihilation of neutrinos into ultralight DM, $g' f' \lesssim 6\times 10^{-8}$. Such small values of couplings do not lead to any significant flux suppression for $E \sim 1$~PeV. But, these values of $g' f'$ can result in energy dependent flavour ratios of astrophysical neutrinos, leading to new prospects in neutrino astronomy at the future neutrino telescopes~\cite{Karmakar:2020yzn}.

\vspace{20pt}

\noindent {\bf Acknowledgements}

\noindent S.K. thanks Najimuddin Khan and Nivedita Ghosh for useful comments. S.R. acknowledges Paolo Gondolo for discussions at the initial phase of this work. 
The authors also thank Vikram Rentala for bringing the constraint from Supernovae cooling to our attention.
The present work is supported by the Department of Science and Technology, India {\it via} SERB grant EMR/2014/001177 and DST-DAAD grant INT/FRG/DAAD/P-22/2018.

\vfill

\appendix 

\section{Cross-section of neutrino-DM interaction}
\subsection{Kinematics}
\label{appendixA1}

We consider the process of neutrinos scattering off DM particles. 
If the incoming neutrino has an energy $E_1$, the energy of the recoiled neutrino is~\cite{JD}, 
\bea 
\label{A1}
E_{3} & = & \frac{E_1 + m_{\text{DM}}}{2} \Big(1 + \frac{m_{\nu}^2 - m_{\text{DM}}^2}{s} \Big) \nn \\
&&+ \frac{\sqrt{E_1^2 - m_{\n}^2}}{2} \Big[ \Big( 1 - \frac{(m_{\n} + m_{\text{DM}})^2}{s} \Big) \Big(  1 - \frac{(m_{\n} - m_{\text{DM}})^2}{s} \Big) \Big]^{1/2} \cos \theta, \nn
\eea 
where $\theta$ is the scattering angle of the neutrino. The relevant Mandelstam variables are, 
\bea
s &=& (p_1^{\m} + p_2^{\m})^2 = m_{\n}^2 + m_{\text{DM}}^2 + 2 E_1 m_{\text{DM}}, \nn\\
t &=& (p_1^{\m} - p_3^{\m})^2 = 2 m_{\n}^2 + 2 (E_1 E_3 - p_1 p_3 \cos \theta) \sim  2 m_{\n}^2 + 2 E_1 E_3 (1 - \cos \theta). \nn
\eea
The energies of incoming neutrinos are such that, $E_1 \sim p_1$ holds well.
The scattering angle $\theta$ in the centre-of-momentum frame can take all values between 0 to $\pi$, whereas that is the case in the laboratory frame only when $m_{\n} < m_{\text{DM}}$. When $m_{\n} > m_{\text{DM}}$, there exists an upper bound on the scattering angle in the laboratory frame, $\theta_{max} \sim m_{\text{DM}}/m_{\n}$. 

The differential cross-section in the laboratory frame is given by~\cite{PBPal}:
\bea
\frac{d\sigma}{d \Omega} = \frac{1}{64 \pi^2 m_{\text{DM}} p_1} \frac{p_3^2}{p_3(E_1+m_{\text{DM}}) - p_1 E_3 \cos \theta}  \sum_{spin} |\mathcal{M}|^2,
\eea
where $d \Omega = \sin \theta d \theta d \phi$.

\subsection{Amplitudes of various renormalisable neutrino-DM interactions}
\label{appendixA2}
$\bullet$~{\bf Fermion-mediated process}

With the renormalisable interaction presented in eq.~(\ref{fermren}), one obtains the amplitude square for the scattering of high energy neutrinos off DM as, 
\bea
\sum_{spin}|\mathcal{M}|^{2}= C_{L}^{4}\frac{(m_{\nu}^{2}-m_{\text{DM}}^{2})(p_{1}.p_{3})-2(m_{\nu}^{2}-p_{2}.p_{3})(p_{1}.p_{2})}{(u-m_{F}^{2})^{2}}.
\eea
Here, $p_{1}, p_2, p_3$ and $p_4$ are the four-momenta of the incoming neutrino, incoming DM, outgoing neutrino and outgoing DM respectively.

\vspace{10pt}

$\bullet$~{\bf Scalar-mediated process}

The amplitude squared for a scalar-mediated process governed by neutrino-DM interaction given by eq.~(\ref{eq}) reads:
\bea
\sum_{spin}|\mathcal{M}|^{2}= g_{\Delta}^{2}f_{l}^{2}\frac{(p_{1}.p_{3}-m_{\nu}^{2})}{(t-m_{\Delta}^{2})^{2}}.
\eea

The neutrinos are Majorana particles in this case and $g_{\Delta}$ has a mass dimension of unity.

\vspace{10pt}

$\bullet$~{\bf Vector-mediated process}

The square of the amplitude for a vector-mediated process described by eq.~(\ref{vecren}) is given as:
\bea
\sum_{spin}|\mathcal{M}|^{2}=2 g'^{2}f'^{2}
\frac{(p_{2}.p_{1}+p_{4}.p_{1})^{2}-(p_{1}.p_{3})(m_{\text{DM}}^{2}+p_{2}.p_{4})}{(t-m_{Z'}^{2})^{2}}.
\eea

\section{Anomaly cancellation for vector-mediated scalar DM model}
\label{appendix:anomaly}
The charges of the SM and exotic fermions are arranged in such a way that they cancel the ABJ anomalies pertaining to the triangular diagram with gauge bosons as external lines and fermions running in the loop. Such conditions are read as:
\bea
\text{Tr}[\gamma^{5}t^{a} \{t^b,t^c\}]=0,
\eea 
where $t^{a},t^{b},t^{c}$ correspond to the generators of the corresponding gauge group and the trace is taken over all fermions. In an anomaly-free theory, the sum of such terms for all fermions for a certain set of gauge bosons identically  vanishes. 
Here the gauge symmetry under consideration is $SU(3)_{c} \times SU(2)_{L} \times U(1)_{Y} \times U(1)'$ where $U(1)'$ represents the new gauge symmetry. 
In our case, third generation leptons,\ie $L_{\tau}$ and $\tau_R$ are charged under $U(1)'$. 
Thus, a full family of additional chiral fermions, namely $Q_{4}, u_{4R}, d_{4R}, L_{4}$ and $l_{4R}$ are needed in order  to cancel anomalies.
 As the new fermions are an exact replica of one generation of SM fermions, the anomalies involving only SM gauge currents, namely  $U(1)_{Y}^3$, $U(1)_{Y} SU(2)_L^2$, $U(1)_{Y} SU(3)_c^2$ and $U(1)_{Y}$(Gravity)$^2$ are automatically satisfied~\cite{Frampton:1999xi}. 
Still we need to take care of the chiral anomalies involving $U(1)'$ which lead to the following conditions~\cite{Ellis:2017tkh,Carena:2004xs}:
\bea
U(1)'SU(3)_c^2:\, && \text{Tr}[Y' \{\sigma^b,\sigma^c\}]=0 \implies 3(2 Y'_{Q_4} - Y'_{u_{4R}} - Y'_{d_{4R}}) = 0,\nn\\
U(1)'SU(2)_L^2:\, && \text{Tr}[Y' \{\sigma^b,\sigma^c\}]=0 \implies  Y'_{L_{\tau}}+  Y'_{L_4}=0,\nn\\
U(1)'^2 U(1)_Y:\, && \text{Tr}[Y'^2 Y]=0 \implies  Y^{\prime 2}_{L_{\tau}}+ Y_{L_4}^{\prime 2} -  Y^{\prime 2}_{\tau_R} -  Y^{\prime 2}_{l_{4R}}=0,\nn\\
U(1)_{Y}^2 U(1)':\, && \text{Tr}[Y^2 Y']=0 \implies   Y'_{L_{\tau}} + Y'_{L_4} - 2 Y'_{\tau_R}  - 2 Y'_{l_{4R}}=0,\nn\\
U(1)'^3:\, && \text{Tr}[Y'^3]=0 \implies  2 Y'^{3}_{L_{\tau}} + 2 Y'^{3}_{L_4} - Y'^{3}_{\tau_R} -  Y'^{3}_{l_{4R}}=0,\nn\\
\text{Gauge-gravity}:\, && \text{Tr}[Y'] =0 \implies  2 Y'_{L_{\tau}} + 2 Y'_{L_4} - Y'_{\tau_R} - Y'_{l_{4R}}=0.
\label{anomaly}
\eea
 While expanding the trace in above relations, an additional $(-)$ sign for the left-handed fermions is implied. 
Here, $Y_{i}^{\prime}$ stands for the $U(1)'$ hypercharge of the species $i$, where $i \equiv L_{\tau}, \tau_R, L_4, l_{4R}$. 
As the exotic quarks are uncharged under $U(1)'$, the first condition of eqs.~(\ref{anomaly}) satisfies.
The SM Higgs transforms trivially under $U(1)'$ in order to keep the Yukawa Lagrangian for quarks and the first two generations of leptons $U(1)'$-invariant. 
Thus, in order to make the Yukawa term involving $\tau$ gauge-invariant, one must put $Y'_{\tau_R} = Y'_{L_{\tau}}$, which serves as another condition along with eqs.~(\ref{anomaly}). 
Thus  the $U(1)'$ hypercharges of the respective fields can be determined from eqs.~(\ref{anomaly}) and are mentioned in table~\ref{content}.
 
\vfill

\pagebreak

\section{Summary of neutrino-DM interactions for scalar DM}
\label{summarytable}
The key constraints on the effective and renormalisable interactions for light DM are summarised in table~IV and V. 

\begin{table}[h]
\begin{center}
\begin{tabular}{|P{2cm} |P{4.3cm}|P{8.5cm}|P{2.0cm}| }
 \hline
 Topology &  Interaction &  Constraints &  Remarks \\
 \hline
 I\,1   & {\footnotesize $\frac{c_{l}^{(1)}}{\Lambda^2}$} $(\bar{\nu} i \slashed{\partial} \nu) (\Phi^* \Phi)$    &{\scriptsize \textcolor{Blue}{$c_{l}^{(1)}/\Lambda^2$} $\lsim 8.8\times 10^{-3}$~GeV$^{-2}$, \textcolor{Red}{$c_{e}^{(1)}/\Lambda^2$} $\lsim 1.0 \times 10^{-4}$~GeV$^{-2},$
 
  \textcolor{ForestGreen}{$c_{\m}^{(1)}/\Lambda^2$} $\lsim 6.0 \times 10^{-3}$~GeV$^{-2}$, \textcolor{Mahogany}{$c_{\tau}^{(1)}/\Lambda^2$} $\lsim 6.2 \times 10^{-3}$~GeV$^{-2}$ }
 & {\footnotesize disfavoured}\\ 
\hline
 I\,2   & {\footnotesize  $\frac{c_{l}^{(2)}}{\Lambda^2} (\bar{\nu} \gamma^\mu \nu)(\Phi^* \partial_\mu \Phi$ 
 
 $- \Phi \partial_\mu \Phi^*)$}    & {\scriptsize \textcolor{Blue}{$c_{l}^{(2)}/\Lambda^2$} $\lsim 1.8 \times 10^{-2}$~GeV$^{-2}$, 
 \textcolor{Red}{$c^{(2)}_{e}/\Lambda^{2}$} $\lesssim 2.6 \times 10^{-5}$~GeV$^{-2}$, \textcolor{ForestGreen}{$c_{\m}^{(1)}/\Lambda^2$} $\lsim 1.2 \times 10^{-2}$~GeV$^{-2}$, \textcolor{Mahogany}{$c_{\tau}^{(1)}/\Lambda^2$} $\lsim 1.3 \times 10^{-3}$~GeV$^{-2}$} & {\footnotesize disfavoured} \\
 \hline
 I\,3   & {\footnotesize $\frac{c_{l}^{(3)}}{\Lambda} \bar{\nu^{c}} \nu \,\, \Phi^{\star} \Phi$  }  & {\scriptsize \textcolor{Blue}{$c^{(3)}_{l}/{\Lambda}$} $\leq 0.5$ GeV$^{-1}$ } &{\footnotesize  favoured$^a$} \\

  \hline
 I\,4   & {\footnotesize $ \frac{c_{l}^{(4)}}{\Lambda^3}(\bar{\nu^{c}} \sigma^{\m\n} \nu)(\partial_{\m} \Phi^{*} \partial_{\nu} \Phi$
 
 $ - \partial_{\nu} \Phi^{*} \partial_{\mu} \Phi)$  }  &  {\scriptsize  \textcolor{Blue}{$c_{l}^{(4)}/\Lambda^3$} $\lesssim 2.0 \times 10^{-3}$~GeV$^{-3}$} & {\footnotesize disfavoured}\\

  \hline
 I\,5   & {\footnotesize $  \frac{c_{l}^{(5)}}{\Lambda^3} \partial^{\m}(\bar{\nu^{c}}  \nu) \partial_{\m} (\Phi^{*}  \Phi)$ }   & {\scriptsize  \textcolor{Blue}{$c_{l}^{(5)}/\Lambda^3$} $\lesssim 7.5 \times 10^{-4}$~GeV$^{-3}$} & {\footnotesize disfavoured}\\

 \hline
 I\,6   & {\footnotesize $ \frac{c_{l}^{(6)}}{\Lambda^4}(\bar{\nu}  \partial^{\m} \gamma^{\nu} \nu)(\partial_{\m} \Phi^{*} \partial_{\nu} \Phi$
 
 $- \partial_{\nu} \Phi^{*} \partial_{\mu} \Phi)$} & {\scriptsize \textcolor{Blue}{$c_{l}^{(6)}/\Lambda^4$} $\lesssim 2.5 \times 10^{-5}$~GeV$^{-4}$, 
 \textcolor{Red}{$c_{e}^{(6)}/\Lambda^4$} $\lesssim 1.2 \times 10^{-6}$~GeV$^{-4}$,
 \textcolor{ForestGreen}{$c_{\m}^{(6)}/\Lambda^4$} $\sim$ \textcolor{Mahogany}{$c_{\tau}^{(6)}/\Lambda^4$} $\lesssim 10^{-5}$~GeV$^{-4}$} & {\footnotesize disfavoured}\\

\hline
 II\,1   & {\footnotesize $ \frac{c^{(7)}_{l}}{\Lambda^{2}}(\partial^{\mu}\Phi^{*} \partial^{\nu}\Phi$ 
 
 $- \partial^{\nu}\Phi^{*} \partial^{\mu}\Phi )Z'_{\mu \nu}$

$ + f_i \bar{\n}_i \gamma^{\mu} P_L \n_{i} Z'_{\mu}$} & {\scriptsize \textcolor{Blue}{$f_l c_{l}^{(7)}/\Lambda^2$} $\lesssim 4.2 \times 10^{-2}$~GeV$^{-2}$, 
\textcolor{Red}{$f_e c_{e}^{(7)}/\Lambda^2$} $\lesssim  1.9 \times 10^{-5}$~GeV$^{-2}$,
\hspace{25pt} \textcolor{ForestGreen}{$f_{\m} c_{\m}^{(7)}/\Lambda^2$} $\sim$ \textcolor{Mahogany}{$f_{\tau} c_{\tau}^{(7)}/\Lambda^2$} $\lesssim  8.1 \times 10^{-3}$~GeV$^{-2}$,\hspace{56pt}
  [\textcolor{ProcessBlue}{$f_e , f_{\m}$}, \textcolor{Mahogany}{$f_{\tau}$}] $\lesssim [10^{-5}, 10^{-6}, 0.02] $ for  $m_{Z'} \sim 10$~MeV}  & {\footnotesize disfavoured}\\

\hline
 II\,2   & {\footnotesize $\frac{c^{(8)}_{l}}{\Lambda} \partial^\mu |\Phi|^2 \partial_\mu \Delta +f_{l} \bar{\nu^{c}}  \nu \Delta$} &  {\scriptsize $ m_{\n} \sim f_{l} v_{\Delta}  \lesssim 0.1$~eV, $m_{\Delta}
\gtrsim 150~$GeV} &  {\footnotesize disfavoured}\\

\hline
 III   & {\footnotesize $C_{1}(\Phi^* \partial_\mu \Phi- \Phi \partial_\mu \Phi^*)Z'^\mu $
 
 $+  \frac{c^{(9)}_{l} }{\Lambda} (\bar{\nu^{c}} \sigma_{\mu \nu} P_L \n)Z'^{\mu \nu}$ }&  {\scriptsize \textcolor{Blue}{$C_1 c_{l}^{(9)}/\Lambda$} $\lesssim 3.8\times 10^{-3}$~GeV$^{-1} $ for  $m_{Z'} \sim 10$~MeV} &   {\footnotesize favoured$^b$ } \\

\hline
 IV   & {\footnotesize $\frac{c^{(10)}_{l}}{\Lambda^2}\bar{L} F_R \Phi |H|^2 + C_{L} \bar{L} F_R \Phi $ } & 
{\footnotesize Same as in fermion case in table~V}
 & {\footnotesize disfavoured }\\
 \hline
\end{tabular}

\caption{Summary of neutrino-DM effective interactions. 
 $c_{l}$ and $c_{e,\m,\tau}$ represent the coefficients of interactions for the gauge non-invariant and gauge-invariant forms respectively. 
The color coding for the constraints is: \textcolor{Blue}{$Z \rightarrow inv$}, \textcolor{Red}{LEP monophoton$+ \slashed{E}_{T}$}, \textcolor{ForestGreen}{$Z \rightarrow {\mu}^{+} {\mu}^{-}$}, \textcolor{Mahogany}{$Z \rightarrow {\tau}^{+} {\tau}^{-}$} and \textcolor{ProcessBlue}{$(g-2)_{e,\mu}$}.
We also remark whether the interactions are favoured in context of the $1\%$ flux suppression criteria as mentioned earlier.}
\end{center}
\label{column1}
\end{table}

\noindent $^a$ \footnotesize{disfavoured if realised with a $SU(2)_L$ triplet scalar.}

\noindent $^b$  \footnotesize{favoured if $0.08~\text{eV} \lesssim m_{\text{DM}} \lesssim 0.5$ eV for $m_{Z'} \sim 10$~MeV and $E_{\n} \sim 1$~PeV.}

\begin{table}[h]
\begin{center}
\begin{tabular}{|P{1.5cm} |P{4.5cm}|P{7.8cm}|P{2cm}| }
 \hline
Mediator& Interaction & Constraints &  Remarks \\
 \hline
  {\footnotesize Fermion} &  {\footnotesize $(C_L \bar{L} F_R  + C_R \bar{l}_R F_L)\Phi + h.c.$} & $m_F \gtrsim 100$~GeV, $m_{\text{DM}} \gsim 10^{-21}$~eV, 
  
  \textcolor{ProcessBlue}{$C_{L}C_{R}$} $\lesssim  \{2.5, 0.5\} \times 10^{-5}$  for $e$ and $\m$ & {\footnotesize disfavoured} \\
\hline
{\footnotesize Scalar}  &  {\footnotesize $f_l \bar{L}^{c} L \Delta   + g_\Delta \Phi^* \Phi |\Delta|^2$} & $m_{\nu} \sim f_{l} v_{\Delta} \lesssim 0.1$~eV, $g_{\Delta} \sim  v_{\Delta}^{2}/m_{\text{DM}}^2$ & {\footnotesize disfavoured} \\
\hline
 {\footnotesize Vector} &  
 
 {\footnotesize $f'_l \bar{L} \gamma^{\mu} P_L L Z'_{\mu} + ig^{\prime}(\Phi^{*} \partial^{\mu} \Phi$ 
 
  $-\Phi \partial^{\mu} \Phi^{*}) Z'_{\mu}$} 
  &  {\footnotesize [\textcolor{ProcessBlue}{$f_e, f_{\m}$}, \textcolor{Mahogany}{$f_{\tau}$}] $\lesssim [10^{-5}, 10^{-6}, 0.02] $ for  $m_{Z'} \sim 10$~MeV}
 & {\footnotesize favoured  only for $\nu_{\tau}$} \\
\hline
\end{tabular}
\caption{Summary of renormalisable neutrino-DM interactions. Color coding is the same as in table~IV. }
\end{center}
\label{column2}
\end{table}

{\normalsize For DM with higher masses the cosmological constraints,\ie relic density, collisional damping and $N_{\text{eff}}$ ensure that the above-mentioned interactions do not lead to any significant flux suppression. This has been discussed in Sec.~\ref{EFT} and ~\ref{CTR}.}

\end{document}